\documentclass[aip,jcp,reprint]{revtex4-1}

\draft 

\usepackage[latin1]{inputenc}
\usepackage{amsmath}
\usepackage{amsfonts}
\usepackage{amssymb}
\usepackage{graphicx}
\usepackage{bm}
\usepackage[titletoc]{appendix}
\usepackage{verbatim}

\usepackage[usenames, dvipsnames]{color}

\DeclareFontFamily{U}{wncy}{}
\DeclareFontShape{U}{wncy}{m}{n}{<->wncyr10}{}
\DeclareSymbolFont{mcy}{U}{wncy}{m}{n}
\DeclareMathSymbol{\Sh}{\mathord}{mcy}{"58} 

\newcommand{\lmax}{\lambda_{\mathrm{max}}}
\newcommand{\lmin}{\lambda_{\mathrm{min}}}

\newcommand{\phat}{\hat{\textbf{p}}}
\newcommand{\Pphat}{\mathcal{P}_{\widehat{\textbf{p}}}}
\newcommand{\Oseen}{\bm{\mathcal{O}}}
\newcommand{\Pb}{\mathcal{P}_{\textbf{b}|\lambda}}
\newcommand{\Plambda}{\mathcal{P}_{\lambda}}
\newcommand{\Pacc}{P_{\mathrm{acc}}}
\newcommand{\Pasym}{\Pacc^{\mathrm{asym}}}
\newcommand{\delx}{\nabla_{\!\textbf{x}}}
\newcommand{\DUest}{\Delta U_{\mathrm{est}}}
\newcommand{\Nlayers}{N_{\mathrm{layers}}}

\begin{document}


\title{
Wavelet Monte Carlo dynamics: a new algorithm for simulating the hydrodynamics of interacting Brownian particles
} 



\author{Oliver T. Dyer}
\author{Robin C. Ball}
\affiliation{Department of Physics, University of Warwick, Coventry, CV4 7AL, UK}


\date{\today}

\begin{abstract}
We develop a new algorithm for the Brownian dynamics of soft matter systems that evolves time by spatially correlated Monte Carlo moves.
The algorithm uses vector wavelets as its basic moves and produces hydrodynamics in the low Reynolds number regime propagated according to the Oseen tensor.
When small moves are removed the correlations closely approximate the Rotne-Prager tensor, itself widely used to correct for deficiencies in Oseen.
We also include plane wave moves to provide the longest range correlations, which we detail for both infinite and periodic systems.
The computational cost of the algorithm scales competitively with the number of particles simulated, $N$, scaling as $N\ln N$ in homogeneous systems and as $N$ in dilute systems.
In comparisons to established lattice Boltzmann and Brownian dynamics algorithms the wavelet method was found to be only a factor of order 1 times more expensive than the cheaper lattice Boltzmann algorithm in marginally semi-dilute simulations, while it is significantly faster than both algorithms at large $N$ in dilute simulations.
We also validate the algorithm by checking it reproduces the correct dynamics and equilibrium properties of simple single polymer systems, as well as verifying the effect of periodicity on the mobility tensor.
\end{abstract}


\maketitle 



%
%

%

\section{Introduction}
\label{sec: Introdunction}

Brownian dynamics (BD) algorithms aim to simplify soft matter simulations by replacing the large number of degrees of freedom in the solvent with known hydrodynamic interactions (HIs), which simply need to be calculated between the $N$ particles of interest at each time step.\cite{Ermak1978BD}

This compares to explicit solvent algorithms that follow solvent molecules with some level of coarse graining, despite not being interested in them directly, to let them mediate viscosity and HIs through local interactions.
Such methods include  molecular dynamics (MD),\cite{Kremer1990MD} dissipative particle dynamics (DPD), \cite{Hoogerbrugge1992DPD} lattice Boltzmann (LB) \cite{Ladd1994LBtheory,Ladd1994LBnumerics,Pham2009LBvsBD} and multi-particle collision dynamics (MPCD) \cite{Malevanets1999MPCD} algorithms.
The computational cost of these methods, or time taken to run a simulation evolving a system by some physical amount of time, scales linearly with the total number of particles. 
This includes particles in the solvent as well as those of interest.
For systems of fixed concentration this leads to the cost scaling as $N$, \cite{Jain2012BD} though the overhead cost of moving the solvent molecules limits the feasible system size, especially as the systems become more dilute.
When considering a non-periodic system the scaling rises.
The important example of single polymer chains leads to $N^{3\nu}$, with $\nu$ the Flory exponent, in order to fit the whole chain inside the simulation box. \cite{Pham2009LBvsBD}

Despite reducing the number of degrees of freedom significantly, the cost of conventional BD algorithms, which is dominated by the decomposition of the mobility tensor, limits their ability to simulate large $N$ systems.
Fixman's algorithm is well known to cut the scaling of this decomposition down to $N^{2.25}$ from the naive $N^{3}$, \cite{Fixman1986BD} and several methods have since been put forward to reduce the scaling further. \cite{Jain2012BD,Saadat2014BD} 
It has even been reduced to or near $N\ln N$ in some cases, although these approaches are only valid for bounded systems,\cite{HernandezOrtiz2006BD} or introduce errors to allow more efficient computation via particle mesh Ewald techniques\cite{Sierou2001BDFFT,Banchio2003BDFFT} or sparse arrays.\cite{Saadat2015BD}

In this work we present an entirely different approach to BD, using a Monte Carlo (MC) algorithm to bypass explicit calculations with the mobility tensor altogether.
To date, MC methods in the field have been used primarily to study equilibrium properties since they have not accounted for HIs.
A variety of different particle movement schemes have been used, including simple methods moving individual particles\cite{Nedelcu2009MC} and evolving the system with torsional rotations of bonds in polymers. \cite{Dodd1993MC, Betancourt2005MC, Zamuner2015MC}
So called bridging moves have also been introduced to handle branching polymers,\cite{Uhlherr2001Bridging, Karayiannis2003MC} and more recently `event-chain' algorithms have been introduced to handle hard sphere particles.\cite{Bernard2009EC,Kampmann2015EC}

An important bridge between the MC methods above and our hydrodynamically coupled method below lies in the work of Maggs.\cite{Maggs2006MC}
Maggs introduced spatially extended correlated MC moves which he tuned to maximise the equilibration rate of a simple fluid system.
He did not target HIs per se, but he was led to motion of the same scaling with arbitrary moves. 
We focus on moves described by wavelets, a class of function that has seen use in many areas of physics, particularly for signal processing because they form an (over) complete, localised basis that allows temporal changes in a signal to be identified. \cite{GrossmannMorlet1984,vandenBerg2004Wavelets} 
The development of wavelet theory is catalogued in Ref.~\citenum{FundamentalPapersWavelets}.
For this work, their role as basis functions enables us to re-express the mobility tensor in a form readily transferable to a MC simulation and hence we call the method `Wavelet Monte Carlo dynamics' (WMCD).
We will show that this approach leads to a cost scaling that is at worst $N\ln N$ per physical unit of time with a very competitive prefactor and no assumptions on the system beyond those already in basic BD algorithms.

In section \ref{sec: Method Sketch} we begin with sketch calculations to highlight how HIs and the cost scaling arise in WMCD without needing the full details.
Section \ref{sec: Hydro tensors} addresses the background physics and mathematics required for the method before section \ref{sec: Method Description} explains the wavelet method in full.
Section \ref{sec: Fourier Method} describes a necessary modification to include occasional plane wave moves, before results of simple validation simulations are given in section \ref{sec: Validation}.

\section{Note on systems, units and hardware}
\label{sec: Units and potentials}

For the data given in this paper, but not required by the theory, we have simulated polymeric systems in a good solvent. 
The polymers are represented by bead-spring chains with finitely extensible non-linear elastic (FENE) springs and the Weeks-Chandler-Anderson (WCA) potential acting between all particles at positions $\textbf{r}_{i}$:
\begin{eqnarray}
U_{\mathrm{FENE}} & = & -\frac{1}{2}k_{\mathrm{FENE}}R_0^2\,
	\ln \left( 1-(r_{ij}/R_0)^2 \right),
\label{eq: FENE potential} \\
U_{\mathrm{WCA}} & = & 4\epsilon 
	\left( \left(\dfrac{\sigma}{r_{ij}}\right)^{12} - \left(\dfrac{\sigma}{r_{ij}}\right)^{6} + 
		\dfrac{1}{4}\right),
\label{eq: WCA potential}
\end{eqnarray}
for $r_{ij}=|\textbf{r}_{ij}|=|\textbf{r}_{i} - \textbf{r}_{j}|< R_{0}, 2^{1/6}\sigma$ respectively.

We adopt as length and energy units $2^{1/6}\sigma$ and $\epsilon$, and denote non-dimensionalised quantities with a bar.
Therefore, to match physical systems in Ref.s~\citenum{Pham2009LBvsBD} and \citenum{Jain2012BD} for benchmarking performance, we use $\bar{\sigma}=2^{-1/6}$, $\bar{\epsilon}=1$, $\bar{k}_{\mathrm{FENE}}=7\times 2^{2/6}$ and $\bar{R}_{0}=2\times 2^{-1/6}$.
Matching hydrodynamic radii of individual particles, $a$, leads to the minimum wavelet radius (introduced in section \ref{sec: lambda PDF}) $\bar{\lambda}_{\mathrm{min}}=0.700$, while the thermal energy is $\overline{k_{B}T}=1.2$.

We define our unit of time to be the time over which completely isolated or non-interacting particles are expected to have diffused by their own radius:
\begin{equation}
\tau = \pi\eta a^{3}/k_{B}T.
\label{eq: Unit of Time}
\end{equation}
By requiring the same viscosity $\eta$ as used in Ref.s~\citenum{Pham2009LBvsBD} and \citenum{Jain2012BD}, which will only appear implicitly in our algorithm, this unit of time is $4.04$ times smaller than in those papers.

Finally, to match the systems in Ref.~\citenum{Jain2012BD}, whenever a system is called `semi-dilute' it consists of polymer chains of length $N_{b}=10$ beads and a global bead concentration $N/\bar{L}^{3}=0.625$, where $L$ is the side length of our simulation box.

All data were obtained using the gcc compiler with optimisation -O3 on a single CPU on an Intel Core2 Quad CPU Q9400 at 2.66GHz.

\section{Method Sketch and Motivation} 
\label{sec: Method Sketch}

\begin{figure}
\includegraphics{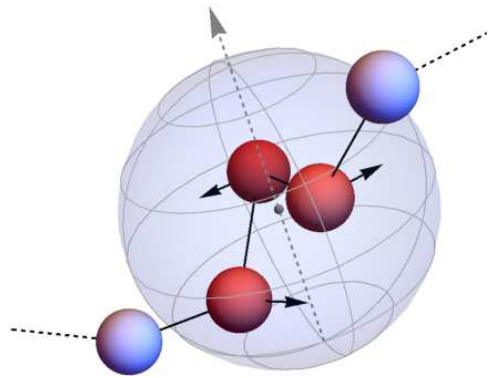}
\centering
\caption{Schematic diagram of a wavelet move on a section of polymer.
The dashed arrow represents $\phat$ and the central dot marks $\textbf{b}$.
The particles enclosed in the wavelet move according to the vector $A_{w}\textbf{w}$, introduced in section \ref{sec: Wavelet representation}, while particles lying outside remain stationary.}
\label{fig: Wavelet diagram}
\end{figure}

We seek a MC algorithm that produces hydrodynamics in a simple and efficient way.
The moves considered are displacements inside spheres of radius $\lambda$ centred on position $\textbf{b}$, chosen from probability density functions (PDFs) $\Plambda$ and $\Pb$ respectively.
Here we have anticipated some $\lambda$-dependence in the distribution for $\textbf{b}$.
The orientation of these moves, $\phat$, is unbiased so that average motion is only induced by non-zero forces weighting the Metropolis test.
An example of such a move is depicted in Fig.~\ref{fig: Wavelet diagram}.

A move will only contribute to correlated motion of particles $i$ and $j$ if it encloses both of them, requiring $2\lambda\geqslant r_{ij}$ and for $\textbf{b}$ to land in a volume of order $\lambda^{d}$ in $d$ dimensions.
The contribution from a given move therefore has the piecewise form
\begin{equation}
\left\langle \delta r_{i}\delta r_{j}\right\rangle \sim\begin{cases}
0 & \textrm{for}\:\, 2\lambda<r_{ij}\\
\Pb \lambda^{d} A_{w}^{2} & \textrm{for}\:\, 2\lambda\geqslant r_{ij}
\end{cases},
\label{eq: drdr sketch}
\end{equation}
where $A_{w}$ is a displacement amplitude that is related to strain by $\varepsilon\propto A_{w}/\lambda$ if, for simplicity in this section, the strain is assumed to be uniform over the move.
In that case the estimate for the change in energy over a move containing $n$ particles is $\DUest \sim n\varepsilon^{2}$, which we are constrained by a Metropolis test to avoid being large compared to $k_{B}T$.
This sets $A_{w} \sim \lambda /\sqrt{n}$.

For $\Pb$ we note that moves with $n=0$ contribute no particle motion and are desirably avoided altogether.
Those with $n\geqslant 1$ can be chosen by first picking a particle at random,
and then choosing $\textbf{b}$ at random within distance $\lambda$.
$\Pb$ is then exactly proportional to $n/(N\lambda^{d})$ since each of the $n$ particles could have led to that centre being chosen. 

Finally we pick $\Plambda \sim \lambda^{-d-1}$, so that integrating Eq.~\eqref{eq: drdr sketch} over radii we have
\begin{equation}
\left\langle \delta r_{i}\delta r_{j}\right\rangle \sim 
	\int\limits_{r_{ij}/2}^{\infty} \!\! d\lambda \lambda^{-d+1}/N
	\propto r_{ij}^{2-d}/N,
\end{equation}
matching the required form for hydrodynamics in $d$ dimensions.

We now know the algorithm produces the desired dynamics so we turn our attention to estimating the computational cost to evolve our system by a physical amount of time.
This will be equal to the cost per move divided by the time evolved per move.
The time factor comes from the previous result, using Brownian motion to set it proportional to $r_{ij}^{2-d}\delta t$ so that we have a time evolution per move going as
\begin{equation}
\delta t \sim 1/N.
\end{equation}

Meanwhile, the cost per move of radius $\lambda$ is of order the expected number of particles involved $n(\lambda)$, so the increment in the computational cost is given by $\delta C=c\int d\lambda\mathcal{P}_{\lambda}n(\lambda)$, with $c$ being the cost per particle.
The homogeneous case has the cost integral diverging logarithmically, leading to 
\begin{equation}
\frac{\delta C}{\delta t}\sim cN\ln N.
\end{equation}
The factor of $N$ here simply reflects the distribution of computational effort across the system, while the $\ln N$ will be seen in section \ref{sec: CPU Cost for Wavelets} to stem from an increase in allowed $\lambda$ values.
It is also shown that for a fractal system, such as a single polymer chain, the logarithm is no longer present.

The method outlined above is therefore very simple, with a favourable cost scaling compared to conventional BD and without the solvent degrees of freedom of explicit solvent methods.
The rest of this paper describes the method in detail for $d=3$.

\section{Mobility Tensors}
\label{sec: Hydro tensors}

We focus on the low Reynolds number limit in which the dynamics of the fluid solvent is governed by the Stokes equations. 
For simplicity we assume we have spherical particles subject to no applied torques, so the dynamics in the system reduces to interrelating the particle velocities, $\textbf{v}_{i}$, to the applied forces $\textbf{F}_{j}$, the latter including interparticle forces. 
In general this is given by
\begin{equation}
\textbf{v}_{i}=\sum_{j}\bm{\mathcal{G}}_{ij} \cdot
	\textbf{F}_{j}+\delta \textbf{v}_{i},
\end{equation}
where $\bm{\mathcal{G}}$ is the mobility tensor and the superposability of the velocity response follows from the linearity of the Stokes equations. 
The contribution from random thermal fluctuations, $\delta \textbf{v}_{i}$, have covariance set by the fluctuation dissipation theorem (FDT), which in the Stokes limit reduces to $\left\langle \delta \textbf{v}_{i}(t)\otimes \delta \textbf{v}_{j}(t')\right\rangle =2k_{B}T\bm{\mathcal{G}}_{ij}\delta(t-t')$.
In practice we need to implement displacements across a small but non-zero time interval $\delta t$, in terms of which this becomes 
\begin{equation}
\left\langle \delta \textbf{r}_{i} \otimes \delta \textbf{r}_{j}\right\rangle =
	2k_{B}T\,\bm{\mathcal{G}}_{ij}\,\delta t.
\label{eq: Correlations general}
\end{equation}

In general the mobility tensor depends on the entire configuration of particle positions and here we have to simplify, following other workers in using the leading dilute limit expressions.
Thus for the self-terms, with $i=j$, we take the Stokes form $\bm{\mathcal{G}}_{ii}=\textbf{I}/6\pi\eta a$, where $a$ is assumed monodisperse for simplicity of exposition. 
For the cross terms we have $\bm{\mathcal{G}}_{ij}(\textbf{r}_{i},\textbf{r}_{j})=\bm{g}(\textbf{r}_{i}-\textbf{r}_{j})$ when the interference of third particles is ignored, and then to leading order in powers of separation $r=|\textbf{r}_{i}-\textbf{r}_{j}|$ we obtain the Oseen tensor,\cite{Doi1988theory}
\begin{equation}
\bm{g}_{\mathrm{Oseen}}(\textbf{r})=
\Oseen (\textbf{r})=\frac{1}{8\pi\eta r}
	\left(\textbf{I}+
	\hat{\textbf{r}}\otimes\hat{\textbf{r}}\right),
\label{eq: Oseen tensor}
\end{equation}
corresponding to particles comoving with the bare solvent response, and as a result $\nabla\cdot \bm{g}_{\mathrm{Oseen}}(\textbf{r})=0$. 

As is well known, the Oseen tensor alone does not assure a mobility matrix with non-negative eigenvalues\cite{Zwanzig1968Oseen_singular} (crucial for the FDT result) so some modification must always be taken at small $r$ to remedy this.
It has become standard in the recent literature to use the Rotne-Prager (RP) form. \cite{RotnePrager1969}
This incorporates the next-leading power of distance, which also turns out to be divergence free:
\begin{equation}
\bm{g}_{\mathrm{RP}}(\textbf{r})=
\Oseen(\textbf{r})+\frac{1}{12\pi\eta r} \left(\frac{a}{r}\right)^{2}
\left(\textbf{I}
	-3\hat{\textbf{r}}\otimes\hat{\textbf{r}}\right),
\label{eq: Rotne Prager tensor}
\end{equation}
for $r\geqslant 2a$, with a branch for $0\leqslant r\leqslant 2a$ given by
\begin{equation}
\bm{g}_{\mathrm{RP}}(\textbf{r})=
\frac{1}{6\pi\eta a} \left(\textbf{I} -\frac{3r}{32a}
\left(3\textbf{I}
	-\hat{\textbf{r}}\otimes\hat{\textbf{r}}\right) \right).
\label{eq: Rotne Prager tensor short range}
\end{equation}

In WMCD we implement the Oseen tensor modified by a cut-off in the wavelet spectrum.
This is positive definite and in section \ref{sec: Effect of lambda+/-} we show that it can lead to a tensor very close to $\bm{g}_{\mathrm{RP}}$.
It should be noted that whilst these modifications
render the total mobility tensor positive definite, they still underrepresent
the reduction in mutual mobility of two particles at close approach,
for which $\hat{\textbf{r}}\cdot \bm{g}\cdot \hat{\textbf{r}}$ should match the self
term $1/6\pi\eta a$ as $r\rightarrow 2a$.

\subsection{Wavelet representation of the Oseen tensor}
\label{sec: Wavelet representation}

Our method is an adaptation of the continuous wavelet transform in
three dimensions, itself based on the identity for the Dirac delta function
\begin{equation}
\delta(\textbf{r}_{i}-\textbf{r}_{j})=
	\mathcal{N}_{\delta}\int\frac{d\lambda}{\lambda}
	\frac{d^{3}\textbf{b}}{\lambda^{3}}\,
	\frac{1}{\lambda^{3}}
	w\left(\frac{\textbf{r}_{i}-\textbf{b}}{\lambda}\right)w\left(\frac{\textbf{r}_{j}-\textbf{b}}{\lambda}\right).
\label{eq: WT of delta function}
\end{equation}
The choice of `mother wavelet' shape is surprisingly arbitrary, with real wavelets constrained only by the formal requirements \cite{vandenBerg2004Wavelets}
\begin{eqnarray}
\int d^{3}\textbf{x} \, w(\textbf{x})^{2} & < & \infty,
\label{eq: wavelet square integrability}\\
\int d^{3}\textbf{x}\, w(\textbf{x}) & = & 0.
\label{eq: wavelet zero average}
\end{eqnarray}  
The normalising front factor is then given by the wavelet Fourier transform $\tilde{w}(\textbf{k})$: $\mathcal{N}_{\delta}^{-1}= (2\pi)^{2} \int d^{3}\textbf{k} k^{-3}|\tilde{w}(\textbf{k})|^{2}$.


For this work it is advantageous to impose additional constraints.
The first is to limit the support of $w$ so that
\begin{equation}
w(\textbf{x})=0 \hphantom{spa} \mathrm{for}\;\, |\textbf{x}|>1.
\label{eq: wavelet finite extent}
\end{equation}
Then it is clear that $w(\,(\textbf{r}-\textbf{b})/\lambda)$ is non-zero only over a region of radius, or `scale', $\lambda$ centred on $\textbf{b}$. 

Next we modify Eq.~\eqref{eq: WT of delta function} in two ways. 
First we make the wavelet a vector valued function, $\textbf{w}(\,(\textbf{r}-\textbf{b})/\lambda,\phat)$, with additional input `polarisation' variable, $\phat$. 
We require these wavelets to be divergence-free,
\begin{equation}
\nabla\cdot\textbf{w}(\textbf{x},\phat)=0,
\label{eq: wavelet divergencelessness}
\end{equation} 
so that $\bm{g}$ inherits this property below.
It is this constraint of the wavelet being transverse which forces us to supply $\phat$.
Our vector wavelet can now be thought of as a flow field in the vicinity of $\textbf{b}$ extending over a lengthscale $\lambda$.

Our second modification is to change the explicit power of $\lambda$ in
Eq.~\eqref{eq: WT of delta function} so that the dimensions match that of
the Oseen tensor. 
This then leads us to consider
\begin{multline}
\bm{g}(\textbf{r}_{i}-\textbf{r}_{j}) = 
\mathcal{N}_{\bm{g}}
	\int \frac{d\lambda}{\lambda}
	\frac{d^{3}\textbf{b}}{\lambda^{3}}
	d^{2}\phat \\ 
	\times\frac{1}{\lambda}\textbf{w}
	\left(\frac{\textbf{r}_{i}-\textbf{b}}{\lambda},\phat\right)
	\otimes \textbf{w}
	\left(\frac{\textbf{r}_{j}-\textbf{b}}{\lambda},\phat\right).
\label{eq: Wavelet representation Oseen}
\end{multline}
In Appendix \ref{sec: Wavelet Derivation} we show that this is exactly equal to $\Oseen(\textbf{r}_{i}-\textbf{r}_{j})$ when $\lambda$ is integrated from 0 to $\infty$ and with an appropriate choice of the constant $\mathcal{N}_{\bm{g}}$.

Eq.~\eqref{eq: Wavelet representation Oseen} is more enlightening when expressed as an expectation value.
In doing so we introduce a wavelet amplitude $A_{w}$, which subsumes all constant factors but may also depend on the wavelet parameters as well.
This dependence is found in section \ref{sec: dU and Aw}, but for now we assume it is known and write
\begin{multline}
\Oseen(\textbf{r}_i - \textbf{r}_j) =  \\
	\left\langle 
		A_{w}\textbf{w}\!
		\left(
			\frac{\textbf{r}_i-\textbf{b}}{\lambda},
			\phat
		\right)\otimes
		A_{w}\textbf{w}\!
		\left(
			\frac{\textbf{r}_j-\textbf{b}}{\lambda},
			\phat
		\right)
	\right\rangle_{\lambda,\textbf{b},\widehat{\textbf{p}}},
\label{eq: Oseen as <ww>}
\end{multline}
where subscripts on the angle brackets indicate quantities averaged over.
These subscripts are left implicit in later results.
Finally, comparison between Eq.s~\eqref{eq: Correlations general} and \eqref{eq: Oseen as <ww>} immediately identifies $A_{w}\textbf{w}\left(\,(\textbf{r}-\textbf{b})/\lambda,\phat\right)$ as the displacement $\delta \textbf{r}$.

\section{Description of the Wavelet Method}
\label{sec: Method Description}

With the expectation value interpretation in Eq.~\eqref{eq: Oseen as <ww>}, we do not need to compute the integral for each particle at every time step, and can instead sample wavelets with appropriate distributions for $\lambda, \textbf{b}$ and $\phat$.
As these moves add up, the correlated motion of particles reproduces that of the Oseen tensor with thermal noise entering via the stochastic move generation.

The details of how this is implemented are given below and in Appendix \ref{sec: Detailed move generation}, but the basic structure of the algorithm follows the process:

\begin{tabular}{l l}
 1: & generate wavelet parameters from their \\ &  distributions, \\\\
 2: & provisionally move particles according to \\ & the resulting wavelet, \\\\
 3: & calculate the energy change, $\Delta U$, caused \\ & by this move and accept or reject it with \\& a Metropolis probability \\ 
\end{tabular}
\begin{equation}
 \Pacc = \mathrm{min}(1,e^{-\Delta U/k_{B}T}).
 \label{eq: Metropolis test}
\end{equation}
The process is then repeated for a desired number of moves.

In the Metropolis test there is no need to include a term for the Jacobian of the move, as per Maggs,\cite{Maggs2006MC} because we will only consider moves with divergenceless flows.

\subsection{Choice of mother wavelet}
\label{sec: Choice of Wavelet}

The restrictions in Eq.s~\eqref{eq: wavelet square integrability}-\eqref{eq: wavelet divergencelessness} still leave a choice for $\textbf{w}$.
In this article we satisfy these conditions by using the form
\begin{equation}
\textbf{w}(\textbf{r},\phat) = \left\lbrace
\begin{array}{l l}
	\phat \times \nabla \phi(\textbf{r})\; &
	\textrm{for}\:\, |\textbf{r}| \leqslant 1 \\
	\textbf{0} & \textrm{for}\:\, |\textbf{r}|>1
\end{array} \right. ,
\label{eq: General wavelet}
\end{equation}
for some scalar function $\phi$.

We also choose to limit ourselves to continuous wavelets so that the strain tensor is finite everywhere.
This is required by neither wavelet theory nor the algorithm, but does simplify some of the analysis.
Further simplification comes by abbreviating the $m^{\mathrm{th}}$ moment of the square of the Fourier transform of $\phi$ to
\begin{equation}
M_{m} \equiv \int\limits_{0}^{\infty} dk \, k^{m} \tilde{\phi}(k)^2.
\label{eq: phi moment}
\end{equation}

For the data in this article we have used the `tapered wavelet', which in spherical polar coordinates $(r,\theta,\varphi)$, and polarisation vector along the $z$-axis ($\theta=0$), is given by
\begin{equation}
\textbf{w}(\textbf{r}, \hat{\textbf{z}}) = r\sin\theta (1-r)\hat{\bm{\varphi}} \hphantom{spa} \mathrm{for}\;\, r\leqslant 1.
\label{eq: Tapered wavelet}
\end{equation}
The associated $\phi$ and its Fourier transform are
\begin{eqnarray}
\phi(r) & = &  \frac{1}{2}r^{2} - \frac{1}{3}r^{3} - \frac{1}{6} \hphantom{spa} \mathrm{for}\;\, r\leqslant 1,
\label{eq: Scalar wavelet position}\\
\tilde{\phi}(k) & = & 4\pi \,k^{-6}
	\left( 5k\sin k - (k^{2}-8)\cos k -8 \right).
\label{eq: Scalar wavelet Fourier}
\end{eqnarray}


\subsection{Distributions of wavelet orientation and centres}
\label{sec: Centring of Wavelets}

With the mother wavelet chosen, we now need to determine the distributions from which to pick the parameters.
For $\phat$ we take an isotropic distribution with $\Pphat = 1/4\pi$ and wavelet amplitude $A_{w}$ independent of $\phat$.

The PDF for $\textbf{b}$ is more involved as we want to avoid spending CPU time on moves that contain no particles and hence don't evolve the system of interest.
To ensure all wavelets contain at least one particle we first pick a particle, all with equal probability, and then choose $\textbf{b}$ uniformly inside a sphere of radius $\lambda$ centred on this particle.

This approach introduces biases that need to be accounted for.
First, the probability of choosing a position inside a volume element is inversely proportional to the volume of the sphere.
Similarly, the chance that any given particle is chosen is inversely proportional to $N$.
Lastly, if the resulting wavelet contains $n$ particles, there must have been $n$ possible ways to have chosen it, and hence the probability is increased by this factor.
All combined, we have the PDF for $\textbf{b}$ as
\begin{equation}
\Pb (\lambda, n(\textbf{b},\lambda)) = \frac{3}{4\pi\lambda^{3}}\frac{n}{N}.
\label{eq: PDF for b}
\end{equation}

\subsection{Choosing the wavelet amplitude}
\label{sec: dU and Aw}

Before the PDF for $\lambda$ can be determined, we need to know the form of $A_{w}$.
To find this we note that if we want the distribution of $\lambda$ to be correctly reflected in the accepted moves, we want $\Pacc$, and hence $\Delta U$, to be as constant as possible over all moves.

An estimate of $\Delta U$ can be made with the strain energy associated with a move
\begin{equation}
\Delta U_{\mathrm{est}} = \frac{1}{2}\mu \frac{3 n}{4\pi\lambda^{3}}
	\int d^{3}\textbf{r}
	\left(\bm{\varepsilon}:\bm{\varepsilon} + 
	\bm{\varepsilon}:\bm{\varepsilon}^{\mathrm{T}}\right),
\label{eq: Wavelet strain energy definition}
\end{equation}
where $\mu$ is a system dependent particle interaction energy expected to be of order $k_{B}T$ in soft-matter systems and leading us to estimate the local shear modulus as $3\mu n/(4\pi \lambda^{3})$.
$\bm{\varepsilon}$ is the move's strain tensor, and $:$ denotes a double dot product.
On the assumption of small displacements we use the infinitesimal strain tensor  $\bm{\varepsilon} = A_{w} \left(\nabla \textbf{w} + (\nabla \textbf{w})^{\mathrm{T}}\right)/2$.

\begin{figure}
\includegraphics{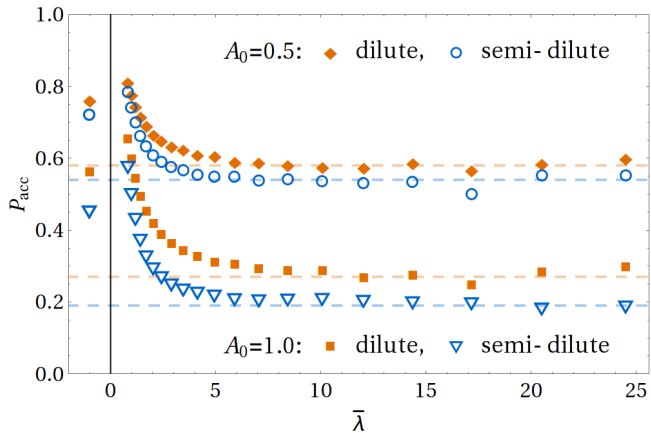}
\centering
\caption{Acceptance probabilities over the spectrum of wavelet radii at different move amplitudes.
The dilute data used an isolated polymer chain with $N=2048$.
The dashed lines are only to emphasize the asymptotic behaviour, while the markers at $\bar{\lambda}=-1$ indicate the average $\Pacc$ over all wavelets.}
\label{fig: wavelet acceptance plot}
\end{figure}

With no preferred direction, $n$ accounts for all location dependence so $\textbf{b}$ and $\phat$ can be left out of the calculation, leaving only the transformation $r \rightarrow r/\lambda$ from the mother wavelet.
In Appendix \ref{sec: Simplifying strain integral} we show how the energy estimate can be written as
\begin{equation}
\DUest = \mu \frac{n}{\lambda^{2}} \frac{A_{w}^{2}}{(2\pi )^{3}} M_{6}
\label{eq: Wavelet strain energy general}
\end{equation}
for any wavelet of the form in Eq.~\eqref{eq: General wavelet}.
For this to be constant over moves we must choose
\begin{equation}
A_{w} (\lambda, n(\textbf{b},\lambda)) = A_{0}\lambda/\sqrt{n},
\label{eq: Wavelet amplitude}
\end{equation}
with $A_{0}$ a dimensionless tuning parameter that scales the maximum rotation angle of the wavelet vector field, and thus must be kept small enough to keep the infinitesimal strain approximation valid.
To maintain simplicity in our results we use $A_{0}$ as a completely free parameter, but we comment that, via Eq.~\eqref{eq: Wavelet strain energy general}, it is fully defined such that
\begin{equation}
\frac{\DUest}{k_{B}T} = \frac{\mu}{k_{B}T} \frac{M_{6}}{(2\pi)^{3}} A_{0}^{2},
\label{eq: A0 definition}
\end{equation}
highlighting the contributions from the choice of mother wavelet ($M_{6}$), the system being simulated ($\mu/k_{B}T$), and the energy scale we are really choosing ($\DUest /k_{B}T$).
The interplay is seen in Fig.~\ref{fig: wavelet acceptance plot}, with both increasing $A_{0}$ and system density, i.e. $\mu$, increasing $\DUest$ and hence decreasing $\Pacc$.

Fig.~\ref{fig: wavelet acceptance plot} also reveals that choosing $A_{w}$ as per Eq.~\eqref{eq: Wavelet amplitude} leads to constant $\Pacc$ only asymptotically, where it limits to a value $\Pasym$.
For small wavelets the strain energy estimate is inaccurate as it requires both particles in an interacting pair to be displaced, which is less likely when $\lambda$ is small, to see a relative displacement from $\nabla \textbf{w}$ rather than $\textbf{w}$ itself.
This results in an overestimate of $\Delta U$ as evidenced by the rise in $\Pacc$.
Ideally one would like to modify the distribution of $\lambda$ in attempted moves to compensate for the bias observed in $\Pacc$, but to do so analytically would be complicated.
A simpler and system independent approach is to keep the form of $A_{w}$ in Eq.~\eqref{eq: Wavelet amplitude}, which already guarantees $\Pacc$ does not drop too low, and use the $\lambda$-recycling scheme described at the end of Appendix \ref{sec: Detailed move generation}.

This scheme is still an imperfect correction for the influence of $\Pacc$ on diffusion rates as it fails to take account of any system dependence on $\textbf{b}$ and $\phat$.
It is therefore advantageous to reduce the variation observed in $\Pacc$, which can be achieved by lowering $A_{0}$ as seen in Fig.~\ref{fig: wavelet acceptance plot}.
Thus increased dynamical fidelity is available in return for the increased computational cost of making smaller displacements per move, and in this respect $A_{0}$ can be viewed as the WMCD analogue of the time step size of other algorithms.
We also anticipate switching from Metropolis to a Glauber test,\cite{Glauber1963} and further to smart MC\cite{Rossky1978BDasMC,Allen1989} will see reductions in the variation in $\Pacc$.

\subsection{Distribution of wavelet radii}
\label{sec: lambda PDF}

With $A_{w}^{2}$, $\mathcal{P}_{\phat}$ and $\Pb$ identified, we can now determine $\Plambda$ from Eq.~\eqref{eq: Wavelet representation Oseen} by requiring 
\begin{equation}
\lambda^{-5}  \propto  
	\Plambda \, \Pb \, \Pphat \, A_{w}^{2}.
\end{equation}
We therefore use the PDF 
\begin{equation}
\Plambda (\lambda) = 
	\mathcal{N}_{\lambda}\lambda^{-4};
	\hphantom{space}
	\mathcal{N}_{\lambda} = 3(\lmin^{-3} - \lmax^{-3})^{-1},
\label{eq: PDF for lambda}
\end{equation}
when normalised between $\lmin$ and $\lmax$.
The effect of using these finite bounds instead of $(0,\infty)$ is discussed in section \ref{sec: Effect of lambda+/-}.
For now we simply comment that a finite $\lmin$ is desirable as it regularises the singularity in the Oseen tensor at $r\rightarrow 0$ and gives us a particle radius via $a=\lmin/\lambda_{a}$, with $\lambda_{a}$ a constant dependent only on the choice of mother wavelet.
$\lmax$ is required when considering a simulation in a finite box of side length $L$, where we do not want wavelets to overlap with their periodic images due to the additional computational effort involved in applying multiple displacements to individual particles.

For a homogeneous system an alternative method can be used in section \ref{sec: Centring of Wavelets} above, whereby $\textbf{b}$ is chosen uniformly across the box so that $\Pb=1/L^{3}$, independent of $\lambda$.
In this case it is sufficient to consider the mean $\left\langle n \right\rangle$, using the global rather than local density.
This leads to $A_{w} = A_{0} L^{3}/(N\sqrt{\lambda})$ and, again, $\Plambda = \mathcal{N}_{\lambda}\lambda^{-4}$.

The rapid decay of $\Plambda$ with either approach means small radius moves dominate.
This is reflected in Fig.~\ref{fig: wavelet acceptance plot} with the mean $\Pacc$ being significantly higher than the asymptote.

\subsection{Time evolution per move}
\label{sec: Understanding Time}

Next we need to know how much physical time passes during a simulation, for which we calculate the expected displacement squared of any given particle in a single move, $\left\langle \delta r_{i}^{2} \right\rangle$.
Using the same approach as in Appendix \ref{sec: Simplifying strain integral} this simplifies to
\begin{equation}
\left\langle \delta r_{i}^{2} \right\rangle = 
	\frac{2 A_{0}^{2}M_{4} }{(2\pi)^{3}N} \mathcal{N}_{\lambda} (\lmin^{-1} - \lmax^{-1}) .
\label{eq: wavelet dr2}
\end{equation}
The simulated time increment in a single move, $\delta\bar{t} =\delta t/\tau$, is then
\begin{equation}
\delta \bar{t} = 
\dfrac{\left\langle \delta r_{i}^{2} \right\rangle }{a^{2}} =
	\dfrac{2 A_{0}^{2} M_{4} \lambda_{a}^{2}}{(2\pi)^{3} N }
	\frac{\mathcal{N}_{\lambda} }{\lmin^{3} }
	\left( 1 - \frac{\lmin}{\lmax} \right).
\label{eq: Time step per wavelet}
\end{equation} 
This is consistent with the claim that $A_{0}$, which is the only free parameter if $\lmax$ is taken as large as the simulation allows, is analogous to the choice of time step.

\subsection{Computational cost}
\label{sec: CPU Cost for Wavelets}

We now come to calculating the computational cost.
Here we consider a system with fractal dimension $d_{f}$, so that the expected number of particles in a move is proportional to $(\lambda/s)^{d_{f}}$, with $s$ the mean separation between near-neighbouring particles.

The cost associated with generating move parameters is small compared to the cost of moving the $n$ particles and calculating each of their energy changes.
We can therefore use an average cost per particle per move, $c$, and multiply this by the expected value of $n$ to find the total cost per move.
Dividing this by the time advance per move in Eq.~\eqref{eq: Time step per wavelet} we have cost per unit dimensionless time
\begin{equation}
\frac{dC}{d\bar{t}} \propto \frac{1}{\delta \bar{t} } 
	\left\langle \Pacc^{-1}\right\rangle c \!\!
	\int\limits_{\lmin}^{\lmax}\!\! d\lambda \,
	\Plambda (\lambda/s)^{d_{f}}.
\label{eq: Cost wavelet general}
\end{equation} 
Here $\left\langle \Pacc^{-1}\right\rangle$, defined as
\begin{equation}
\left\langle \Pacc^{-1}\right\rangle \equiv 
	\int\limits_{\lmin}^{\lmax}\!\! d\lambda \,
	\Plambda \lambda^{d_{f}} / \Pacc(\lambda)
	\Bigg/ \int\limits_{\lmin}^{\lmax}\!\! d\lambda \,
	\Plambda \lambda^{d_{f}},
\end{equation}
handles the additional $\lambda$-dependence coming from our $\lambda$ recycling scheme.
Its value lies between 1 and $1/\Pasym$, itself of order unity, and limits to this upper value as $\lmax\rightarrow\infty$.
Its effect on the estimated scaling is therefore small and we will treat it as a constant.

\subsubsection{Cost of homogeneous systems}
\label{sec: Homogeneous cost}
When the system is homogeneous or in a poor solvent so that $d_{f}=3$, the integral evaluates to a logarithm and $s^{3}=L^{3}/N$.
The cost then goes as
\begin{equation}
\frac{dC_\mathrm{homo}}{d\bar{t}} \propto N^{2} \,
	\dfrac{\left\langle \Pacc^{-1}\right\rangle c}{A_{0}^{2} M_{4} \lambda_{a}^{2}}\!
	\left( \frac{\lmin}{L}\right)^{3}\!
	\dfrac{\ln(\lmax/\lmin)}
		{1-\lmin/\lmax}.
\label{eq: Cost wavelet homogeneous}
\end{equation}
For systems with fixed global density and $\lmax\sim L \sim N^{1/3}$, this reduces to $C_{\mathrm{homo}}\sim N\ln N$ per unit of time.

\begin{figure}
\includegraphics{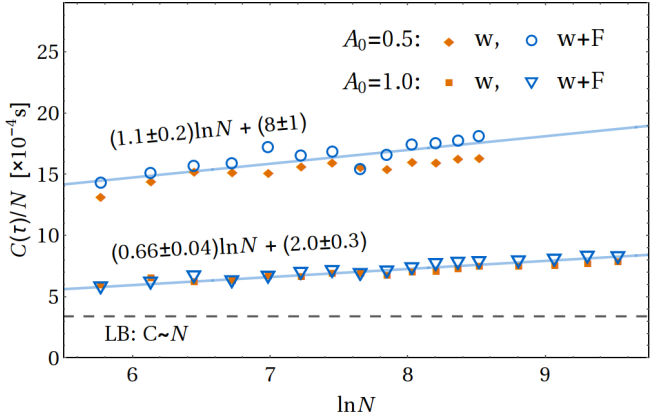}
\centering
\caption{CPU cost per particle to evolve semi-dilute (homogeneous) systems by a single time unit.
The dashed line indicates LB timings from Fig.~8 in Ref.~\citenum{Jain2012BD}, which used identical systems, rescaled to our time unit.
Timings for both pure wavelet (w) and periodic wavelet plus Fourier (w+F) algorithms, the latter described in section \ref{sec: Fourier Method}, are shown for comparison.}
\label{fig: Cost semi-dilute}
\end{figure}

Fig.~\ref{fig: Cost semi-dilute} shows cost timings in semi-dilute homogeneous systems which confirm the $N\ln N$ scaling.
This is true also of the wavelet plus Fourier (w+F) data in this figure, which includes moves described in section \ref{sec: Fourier Method} that only add a minor cost.
The following discussion therefore applies to either pure wavelet or w+F data equally.

Because the $\ln N$ factor originates in $\lmax$, and hence sees only the asymptotic acceptance probability $\Pasym$, the coefficients of the $\ln N$ terms in Fig.~\ref{fig: Cost semi-dilute} vary as $1/(A_{0}^{2}\Pasym)$.
Reading $\Pasym$ from Fig.~\ref{fig: wavelet acceptance plot} predicts a ratio of coefficients of approximately 1.5, agreeing within the error margins in Fig.~\ref{fig: Cost semi-dilute}.

Also in Fig.~\ref{fig: Cost semi-dilute}, the cost for the LB algorithm for identical semi-dilute systems is seen to be a factor of order 1 times faster than the WMCD algorithm, with the exact factor depending on both $A_{0}$ and $N$.
(The equivalent BD costs are several orders of magnitude larger.\cite{Jain2012BD})
Here we note that a fair comparison would use the same hardware for each algorithm, which was not done here, and the ratio between the costs is therefore only a rough indicator of their relative performance.
Moreover our code was far from optimised with particle neighbour lists (the dominant contribution to the cost) being recomputed every move rather than updated only as required, we did not exploit multiple processors, and we compiled with just the standard gcc compiler.
Finally, we expect gains relative to LB in semi-dilute systems with longer chains, which are less dense than the systems used in this comparison and consequently the LB algorithm's explicit solvent would incur an additional cost.

\subsubsection{Cost of fractal systems}
\label{sec: Fractal cost}
When $d_{f}<3$, as in the case for a single polymer chain in good ($d_{f}=1/0.588$) or $\theta$ ($d_{f}=2$) solvents,\cite{Doi1988theory} the cost integrates to
\begin{equation}
\frac{dC_\mathrm{frac}}{d\bar{t}} \propto N\,
	\dfrac{\left\langle \Pacc^{-1}\right\rangle c}{A_{0}^{2} M_{4} \lambda_{a}^{2} } \!
	\left( \frac{\lmin}{s} \right)^{d_{f}}\!
	\dfrac{1-(\lmin/\lmax)^{3-d_{f}}}
		{1-\lmin/\lmax}.
\label{eq: Cost wavelet fractal}
\end{equation}
This time $s$ is taken as the bond length between beads and is set by the potentials between particles.
For $\theta$ solvents the $\lmax$ dependence cancels, while it is only significant for small systems in good solvent.
In either case we are left with $C_{\mathrm{frac}} \sim N$.

\begin{figure}
\includegraphics{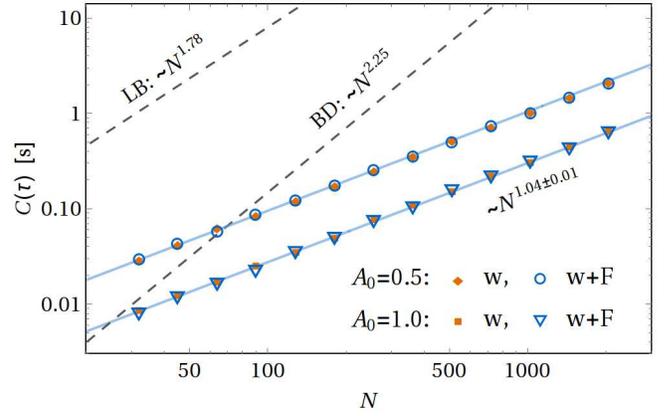}
\centering
\caption{CPU cost to evolve systems with an isolated polymer of length $N$ by a single time unit with the pure wavelet (w) and infinite wavelet plus Fourier (w+F) algorithms.
The systems considered were identical to those in Fig.~11 in Ref.~\citenum{Pham2009LBvsBD}, and the dashed line indicates the BD and LB timings from that plot, rescaled to our time unit.}
\label{fig: Cost dilute}
\end{figure}

This is shown in Fig.~\ref{fig: Cost dilute} where the observed scaling is slightly faster than the theoretical linear scaling in $N$.
We suspect the discrepancy is due to our use of the cell index method,\citep{Allen1989} which introduces small changes in the cost of identifying moving and interacting particles as the chain length increases.
For the same reason, the cost is actually $L$ dependent with the optimal $L$ increasing with $N$.
For the data in Fig.~\ref{fig: Cost dilute} we used $L=50$ for all data points, and hence the costs given are not optimal.

Again, in comparison with other algorithms the differences in hardware should not be forgotten, but in the dilute regime the wavelet algorithm is seen to be much faster than conventional BD for all but the shortest chains, while the LB algorithm is now more expensive because of the many solvent molecules.

\subsection{Effect of $\lmin$ and $\lmax$ on the mobility tensor}
\label{sec: Effect of lambda+/-}

So far the effects of using the finite bounds $\lmin, \lmax$ have only been stated without proof.
To calculate the effects we first take the Fourier transform of the wavelet representation of the Oseen tensor, as per Appendix \ref{sec: Wavelet Derivation}, substitute in the wavelet form in Eq.~\eqref{eq: General wavelet} and impose the finite limits on the $\lambda$ integral.
This results in the Fourier space tensor
\begin{equation}
\tilde{\Oseen}_{w}(\textbf{k};\lmin ,\lmax) = 
	\frac{4\pi A_{0}^{2}}{3 k^{2}} \left(\textbf{I} - \hat{\textbf{k}}\otimes\hat{\textbf{k}}\right)
	\!\! \int\limits_{k\lmin}^{k\lmax} \!\!\! dq\,q^{3}\tilde{\phi}(q)^{2} .
\label{eq: Oseen_w FT}
\end{equation}

The inverse FT then gives the tensor simulated by the algorithm in a form that more clearly shows its characteristics.
The inverse FT with limits $(k\lambda',\infty)$ is found in Appendix \ref{sec: Derivation of O'} to be
\begin{multline}
\Oseen_{w}(\textbf{r};\lambda',\infty) = \frac{A_{0}^{2}}{3\pi r}
	\int\limits_{0}^{\infty} dq\,q^{3}\tilde{\phi}(q)^2 
	\bigg[ \left(\textbf{I} + \hat{\textbf{r}}\otimes\hat{\textbf{r}}\right)
		\mathrm{Si}\,(Q) \\
		+ \left(\textbf{I} - 3\hat{\textbf{r}}\otimes\hat{\textbf{r}}\right)
		\dfrac{\sin Q - Q \cos Q}{Q^2}
	\bigg],
\label{eq: Oseen (min,infty)}
\end{multline}
where $Q = q r/\lambda'$ and $\mathrm{Si}$ is the sine-integral function $\mathrm{Si}\,(Q) = \int_{0}^{Q}\!dt\,\sin t/t$.
The full tensor is then
\begin{equation}
\Oseen_{w}(\textbf{r};\lmin ,\lmax) = 
	\Oseen_{w}(\textbf{r};\lmin ,\infty ) - \Oseen_{w}(\textbf{r}; \lmax ,\infty),
\label{eq: Oseen_w}
\end{equation}
which approximates the Oseen tensor between $\lmin$ and $\lmax$, with regularisation of the singularity and missing correlations for separations larger than $\lmax$.

To calculate the regularisation at $r_{ij}\rightarrow 0$, and hence find the particle hydrodynamic radius, we consider the case where $\lmax\rightarrow\infty$. 
This choice anticipates the modification in section \ref{sec: Fourier Method}.
We then associate Eq.~\eqref{eq: Oseen (min,infty)} with the self- and cross-terms in section \ref{sec: Hydro tensors}, taking the limits $r\rightarrow 0$ and $r\rightarrow\infty$ respectively.
In the former limit, the square brackets in Eq.~\eqref{eq: Oseen (min,infty)} become $(4/3)Q\textbf{I}$, while for the latter they become $(\pi/2)(\textbf{I} + \hat{\textbf{r}}\otimes\hat{\textbf{r}})$.
Since this yields the expected tensor structures it is sufficient to equate the scalar factors and then solve the 2 equations, giving the ratio
\begin{equation}
\lambda_{a} \equiv \frac{\lmin}{a} =  \frac{2}{\pi}\frac{M_{4}}{M_{3}}.
\label{eq: lambda_a}
\end{equation}
Hence and as previously claimed, $\lmin$ is exactly proportional to particle radius so long as the moments $M_{3}$ and $M_{4}$ exist.
For the tapered wavelet, $\lambda_{a} = ((9/8) - \ln 2)^{-1} \approx 2.316$.

\begin{figure}
\includegraphics{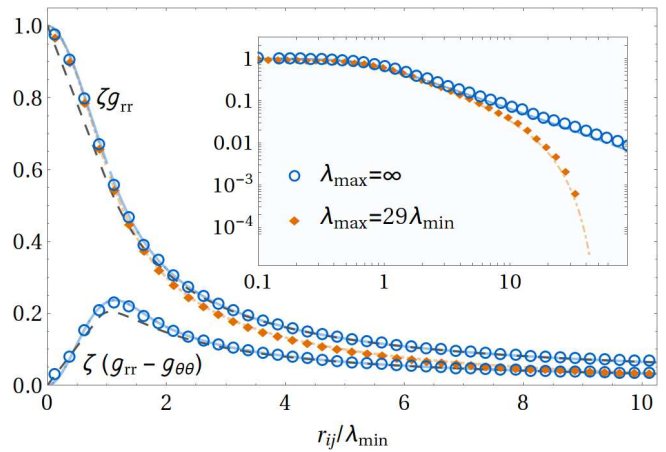}
\centering
\caption{Plots of simulated mobility tensor elements normalised by $\zeta=6\pi\eta a$
with the value of $\lmax$ given in the legend.
Theoretical curves from Eq.~\eqref{eq: Oseen_w} lie underneath the data and curves for the full RP tensor are also shown (dashed) for comparison.
For $\lmax = 29\lmin$ the data for $\zeta (g_{rr}-g_{\theta\theta})$ are not shown: they overlay the $\lmax = \infty$ data over the whole range.
The inset shows $\zeta g_{rr}$ using logarithmic scales.}
\label{fig: Correlation plots}
\end{figure}

Fig.~\ref{fig: Correlation plots} shows both radial and angular elements of the simulated mobility tensor, measured from correlations in particle displacements, plotted alongside curves derived from Eq.~\eqref{eq: Oseen_w}.
The finite $\lmax$ data shows long range deviation from $r_{ij}^{-1}$ behaviour.
Intuitively this must happen as any particles separated by a distance larger than $2\lmax$ cannot possibly have correlated motion due to never being in the same move.
As $\lmax$ increases the long range correlations are more accurately supplied, limiting to the correct $1/r$ behaviour of the $\lmax\rightarrow \infty$ data, obtained using the Fourier moves described in the next section.


Note that the correlations are very close to those of the RP tensor across both of its branches.
As the RP tensor is already an approximation at small $r_{ij}$, modifying the method to explicitly replicate this tensor is not expected to be worthwhile, although we note that it is possible to do so using Fax\'{e}n's laws. \cite{DurlofskyBradyBossis1987}

\section{Adaptation to Include Fourier Moves}
\label{sec: Fourier Method}

We now describe how to modify the algorithm to correct for finite $\lmax$ by adding $\Oseen'(\textbf{r};\lmax)$ back into Eq.~\eqref{eq: Oseen_w}.
This is achieved using essentially the same approach but using plane waves as our moves instead of wavelets.

Although it did not matter for wavelets because $\lmax$ ensured periodic images would not overlap, whether we are considering an infinite or periodic system is now important, and the corresponding tensors will be denoted $\Oseen_{F}^{\infty}(\textbf{r};\lmax)$ and $\Oseen_{F}^{P}(\textbf{r};\lmax)$.
Similarly, other boundary condition dependent quantities will be distinguished with these superscripts, and equations that apply for either will leave the superscript off of these quantities.

We start by writing these tensors in the plane wave basis, which is none other than the Fourier transform, so we have
\begin{equation}
\tilde{\Oseen}_{F}^{\infty}(\textbf{k};\lmax) = 
	\tilde{\Oseen}_{w}(\textbf{k};\lmax ,\infty),
\label{eq: Oseen_F^inf FT}
\end{equation}
which can be read directly from Eq.~\eqref{eq: Oseen_w FT}.

For the periodic case only the commensurate wavevectors are represented, leading to
\begin{eqnarray}
\tilde{\Oseen}_{F}^{P}(\textbf{k};\lmax) & = & 
		\left( \frac{2\pi}{L}\right)^{3}
	\Sh (\textbf{k})\;
	\tilde{\Oseen}_{w}(\textbf{k};\lmax ,\infty);
\label{eq: Oseen_F^P FT} \\
\Sh (\textbf{k}) & \equiv &
	\sum_{\bm{\ell}} \delta^{3}(\textbf{k}-\textbf{k}_{\bm{\ell}}),
\end{eqnarray}
with $\textbf{k}_{\bm{\ell}}=(2\pi/L)(\ell_{x},\ell_{y},\ell_{z})$ and all $\ell \in \mathbb{Z}$. 

\begin{figure}
\includegraphics{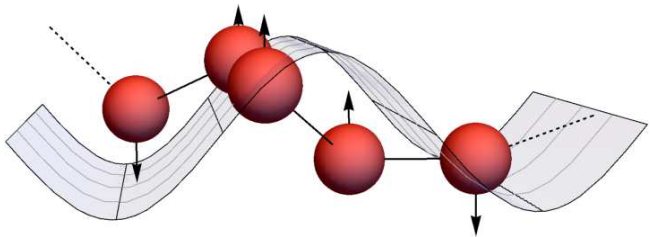}
\centering
\caption{Schematic diagram of how a Fourier move displaces all particles in the system.
The plane wave surface indicates the vector field, which is perpendicular to $\textbf{k}$ and spans the entire system.} 
\label{fig: Fourier diagram}
\end{figure}

The inverse Fourier transforms can be re-expressed in expectation value form by recognising 
$(\textbf{1} - \hat{\textbf{k}}\otimes\hat{\textbf{k}}) = 2\left\langle \hat{\textbf{e}}\otimes\hat{\textbf{e}} \right\rangle_{\widehat{\textbf{e}}}$, with $\hat{\textbf{e}}$ a unit vector perpendicular to $\hat{\textbf{k}}$.
To re-express $e^{i\textbf{k}\cdot\textbf{r}_{ij}}$ we note that $\tilde{\Oseen}_{w}(\textbf{k};\lmax , \infty)$ is even in $\textbf{k}$, so we can replace the complex exponential with
\begin{equation}
\cos (\textbf{k}\cdot\textbf{r}_{ij}) = 
	2 \left\langle \cos(\textbf{k}\cdot\textbf{r}_{i} + \Phi) 
	\cos(\textbf{k}\cdot\textbf{r}_{j} + \Phi)\right\rangle_{\Phi}.
\end{equation}
In both cases the distributions are uniform so that $\mathcal{P}_{\widehat{\textbf{e}}}=\mathcal{P}_{\Phi}=1/2\pi$,
and we finally have
\begin{multline}
 \Oseen_{F}(\textbf{r}_{ij}) = \\
  	\left\langle 
 		A_{F}\cos(\textbf{k}\cdot\textbf{r}_{i} +\Phi)\hat{\textbf{e}}
 		\otimes
 		A_{F}\cos(\textbf{k}\cdot\textbf{r}_{j} +\Phi)\hat{\textbf{e}}
 		\right\rangle_{\Phi,\widehat{\textbf{e}}, \textbf{k}}
\label{eq: Oseen as <FF>}
\end{multline}
with $A_{F}$ filling the same role as $A_{w}$ did for wavelet moves.
We therefore have Fourier moves causing displacements $\delta\textbf{r} = A_{F}\cos(\textbf{k}\cdot\textbf{r} +\Phi)\hat{\textbf{e}}$, as seen diagrammatically in Fig.~\ref{fig: Fourier diagram}, and the infinite and periodic versions differing only in their distributions over $\textbf{k}$.

The rest of this section follows loosely the preceding wavelet section.

\subsection{Choosing plane wave amplitude}
\label{sec: AF and k PDF}

Similarly to section \ref{sec: dU and Aw}, we desire $A_{F}$ to be chosen such that $\Delta U$ is independent of $\textbf{k}$.
Further to this we want the change in energy to be equal to that of wavelet moves.

In a Fourier move the whole simulation box sees a displacement, so the move volume is $L^{3}$ and $n=N$.
Any $L$-dependence will cancel so this approach is still valid for the infinite case.
The strain energy estimate is therefore given by
\begin{eqnarray}
\DUest & = & \frac{1}{2}\mu N L^{-3} A_{F}^{2} 
	k^{2}
	\int\limits_{\mathrm{box}}d^{3}\textbf{r}\sin^{2}
	(\textbf{k}\cdot\textbf{r} + \Phi )
\nonumber \\
& = & \frac{1}{4}\mu N A_{F}^{2} k^{2},
\label{eq: Fourier strain energy}
\end{eqnarray}
setting $A_{F}(k) = 2 k^{-1} \sqrt{\DUest / \mu N}$ with $\DUest /\mu$ taking the same value as set by $A_{0}$ in Eq.~\eqref{eq: A0 definition}.

\begin{figure}
\includegraphics{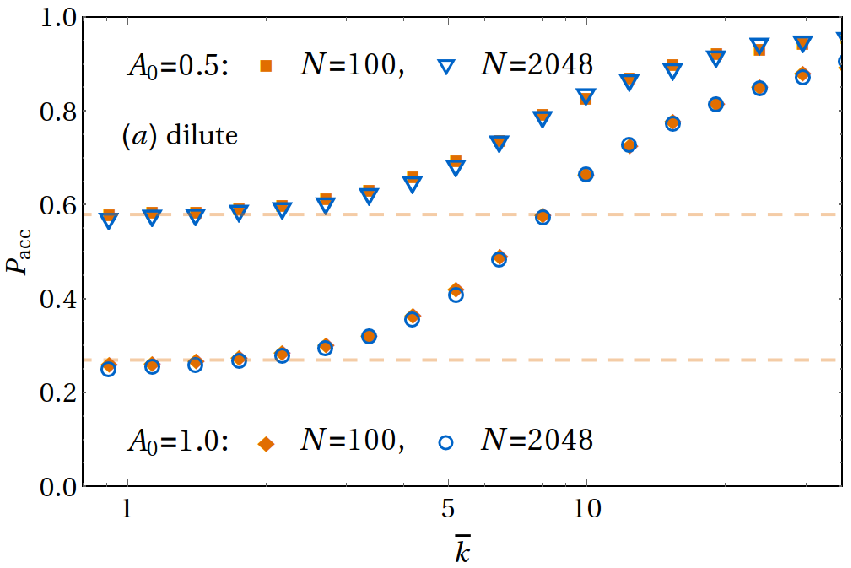}
\includegraphics{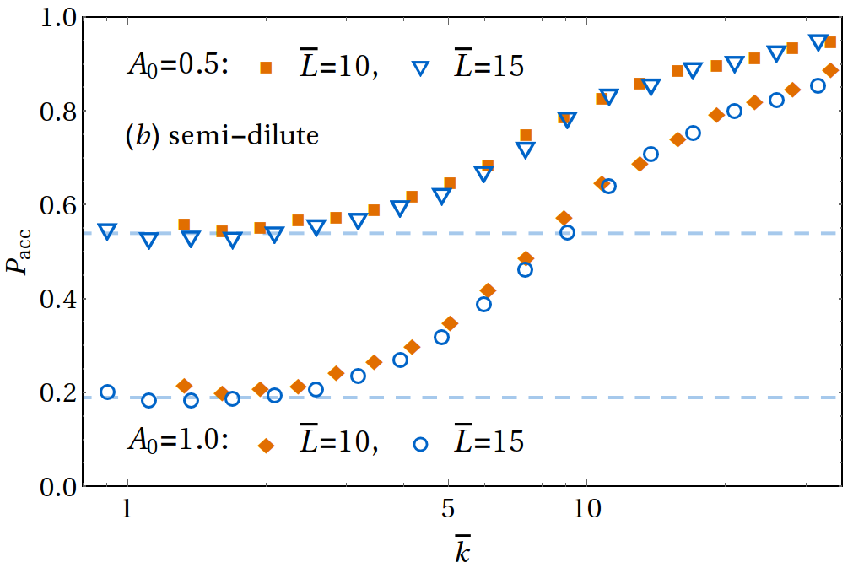}
\centering
\caption{Acceptance probabilities over the spectrum of plane waves different move amplitudes. 
(a): Single chain data using the infinite algorithm. 
(b): Data for semi-dilute systems with the periodic algorithm.
Dashed lines, indicating the asymptotic wavelet values of $\Pacc$ for equivalent systems, are identical to the dashed lines in Fig.~\ref{fig: wavelet acceptance plot}.}
\label{fig: k acceptance prob}
\end{figure}

Fig.~\ref{fig: k acceptance prob} verifies that choosing this form does indeed lead to a constant acceptance probability, at least for small $k$ modes which will end up dominating the distribution.
For these low frequency moves comparison with the wavelet asymptotic values, seen again as dashed lines in Fig.~\ref{fig: k acceptance prob}, confirms equating Eq.s~\eqref{eq: Wavelet strain energy general} and \eqref{eq: Fourier strain energy} correctly equates the actual energy changes in wavelet and Fourier moves.

At large $k$ the wavelength is less than the particle separation and the gradient of the vector field is no longer a good measure of the relative displacements of nearby particles.
Consequently the strain energy is an overestimate, leading to $\Pacc$ rising to 1.
Similarly to how we recycle the radius of failed wavelet moves, and as described at the end of Appendix \ref{sec: Detailed move generation}, we recycle the absolute value of $k$ of Fourier moves to correct for the variation in $\Pacc$.

We now address the divergence of $A_{F}$ at the $k=0$ mode.
This mode is left out in the Ewald sum in other methods to enforce zero net force. \cite{Beenakker1986}
For our algorithm, the $k=0$ mode has no influence on dynamics within the system and by viewing the system in the centre of mass frame we set the displacements associated with this move to zero.
However, because in the periodic case this mode has a non-zero probability of being chosen, as is shown in the next section, we deem these moves be accepted even though they induce no change in the system.
Their automatic acceptance coming from $\Delta U = 0$ accounts for the slight rise in $\Pacc$ in the lowest $k$ bins in Fig.~\ref{fig: k acceptance prob}(b).

\subsection{Distributions of wavevectors}

With the amplitude chosen, we are now able to determine the distribution of wavevectors.
For the infinite case these are isotropically distributed so that $\mathcal{P}_{\widehat{\textbf{k}}}^{\infty}=1/4\pi$, while the radial part picks up $k^{2}$ from $d^{3}\textbf{k}$, giving
\begin{equation}
\mathcal{P}_{k}^{\infty}(k) = \mathcal{N}_{k}^{\infty} k^{2}\!\!\!\!
	\int\limits_{k\lmax}^{\infty} \!\!\!\! dq\,q^{3}\tilde{\phi}(q)^{2}.
\label{eq: PDF of k}
\end{equation}
The normalisation simplifies to
\begin{equation}
\mathcal{N}_{k}^{\infty} = 3\lmax^{3} M_{6}^{-1}.
\label{eq: k normalisation infinite}
\end{equation} 

For the periodic case, $\Sh (\textbf{k})$ turns the PDF into the discrete set of probabilities
\begin{equation}
P_{\textbf{k}}^{P}(\textbf{k}_{\bm{\ell}}) = \mathcal{N}_{\textbf{k}}^{P}\!\!\!\!
	\int\limits_{k_{\bm{\ell}}\lmax}^{\infty}\!\!\!\! dq\,q^{3}\tilde{\phi}(q)^{2},
\label{eq: Prob of k_l}
\end{equation}
where the normalisation, obtained by a sum of this integral over all $\textbf{k}_{\bm{\ell}}$, cannot be  expressed in a simple way.

\begin{figure}
\includegraphics{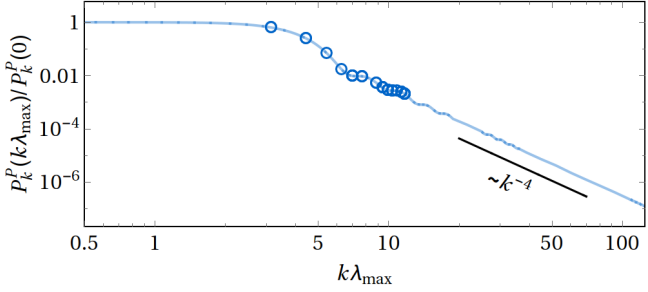}
\caption{Probability of picking Fourier modes in a periodic system in Eq.~\eqref{eq: Prob of k_l}, normalised by the probability of picking the $k=0$ mode.
Markers indicate the lowest discrete modes when $\lmax=L/2$, starting at $k=2\pi/L$.}
\label{fig: k PDF}
\end{figure}

For the tapered wavelet $P_{\textbf{k}}^{P}$ is shown graphically in Fig.~\ref{fig: k PDF}, where the $k^{-4}$ decay in the number of high frequency modes, resulting from wavelets already accounting for the short-range correlations these would be contributing, ensures $\mathcal{N}_{\textbf{k}}^P$ converges.
As the markers in this figure show, when $\lmax$ takes its maximal value of $L/2$ all but the $k=0$ mode experience this decay and low frequency modes therefore dominate the distribution.

\subsection{The probability of making a Fourier move}
\label{sec: Prob of Fourier}

It now only remains to determine how often to make a Fourier move, for which we compare the magnitudes of the tensors generated by the missing wavelets with $\lambda > \lmax$ and the Fourier moves replacing them, \textit{after} conversion to expectation value form.
Surprisingly we find exact equality, i.e.
\begin{equation}
\left\langle \delta \textbf{r}_{i} \otimes \delta \textbf{r}_{j} 
	\right\rangle_{w}^{\lambda >\lmax} = 
	\left\langle \delta \textbf{r}_{i} \otimes \delta \textbf{r}_{j} \right\rangle_{F}^{\infty},
\label{eq: R^inf definition}
\end{equation}
when we might have expected the normalisations to introduce a relative factor.
It turns out the fixing of the strain energies to be equal, which can be viewed as an operator on $\langle \delta \textbf{r}_{i} \otimes \delta \textbf{r}_{j} \rangle$, set this factor to 1.
This leads us to the remarkable result that, on average, a Fourier move evolves time by the exact amount that a missing large wavelet would have done, regardless of the value of $\lmax$.
We therefore need to make a Fourier move with the same probability as picking a wavelet with $\lambda > \lmax$:
\begin{equation}
P_{F}^{\infty} = \frac{\lmin^{3}}{\lmax^{3}}.
\label{eq: P(Fourier) infinite}
\end{equation}

To show that this argument is consistent with the previously found distributions, we explicitly calculate $\left\langle \delta r_{i}^{2} \right\rangle$ for infinite system Fourier moves.
After averaging over space and the trivial $\hat{\textbf{k}}$, $\Phi$ and $\hat{\textbf{e}}$ integrals have been performed this becomes
\begin{eqnarray}
\left\langle \delta r_{i}^{2} \right\rangle_{F}^{\infty} & = & \frac{1}{2}\mathcal{N}_{k}^{\infty}
	\int\limits_{0}^{\infty} dk \, k^{2} A_{F}^{2} \!\!\!
	\int\limits_{k\lmax}^{\infty} \!\!\! dq\, q^{3} \tilde{\phi}(q)^{2} 
\nonumber	\\
    & = & \frac{2 \DUest}{N\mu} \frac{3\lmax^{3}}{\lmax} \frac{M_{4}}{M_{6}},
\label{eq: Fourier dr2}
\end{eqnarray}
which is identical to Eq.~\eqref{eq: wavelet dr2} when $\lmax\rightarrow\infty$ and $\lmin \rightarrow \lmax$, as expected.

In periodic systems the argument is complicated by the estimated energy for wavelets being invalid for $\lambda > L/2$ when they overlap with their own images, and hence the relative factor, $\mathcal{R}$ say, has not been fixed at 1. 
Performing the energy calculation directly is intractable for overlapping wavelets, but the value of $\mathcal{R}$ can be found indirectly by comparing the introduced normalisation factors.
The wavelet factors are unchanged from the infinite case, so it is sufficient to use the ratio of introduced factors between the infinite and periodic Fourier tensors, which gives
\begin{equation}
\mathcal{R} = \left( \frac{L}{2\pi}\right)^{3} 
	\frac{\mathcal{N}_{\textbf{k}}^{P}}{\mathcal{N}_{k}^{\infty}
	\mathcal{N}_{\widehat{\textbf{k}}}^{\infty}}
	= \frac{1}{6\pi^{2}} \left( \frac{L}{\lmax} \right)^{3} 
	M_{6} \mathcal{N}_{\textbf{k}}^{P}.
\end{equation}

To calculate the probability of making a Fourier move we enforce equal time evolution from Fourier and missing wavelet moves after the same number of wavelet moves with $\lambda < \lmax$, leading to
\begin{equation}
P_{F}^{P} = \left( 1 + \mathcal{R} \left[ (\lmax/\lmin)^{3} - 1 \right]\right)^{-1}.
\label{eq: P(Fourier)}
\end{equation}

Finally we calculate the new value of $\delta \bar{t}$, taking both Fourier and wavelet moves into account.
Because $\tau$ is defined for an isolated particle this will use $\left\langle \delta r_{i}^{2} \right\rangle_{F}^{\infty}$ in \textit{both} the periodic and infinite systems.
Since this is identical to the missing contribution of $\lmax <\lambda < \infty$, it is clear we can just take $\lmax\rightarrow \infty$ in Eq.~\eqref{eq: Time step per wavelet} to obtain
\begin{equation}
\delta \bar{t} = \frac{ 6 A_{0}^{2} M_{4}\lambda_{a}^{2}}{(2\pi)^{3} N }.
\label{eq: dt w+F}
\end{equation}

\subsection{Computational cost with Fourier moves}
\label{sec: CPU Cost for Fourier}

Armed with the frequency at which Fourier moves are chosen, we now calculate the computational cost associated with them.
Each plane wave evolves all $N$ particles, each with a cost $c$ as in the wavelet case, by $\delta t \sim N^{-1}$.
For Fourier moves alone then, the cost would scale as $N^{2}$.

Accounting for the presence of wavelet moves, the Fourier contribution to the cost is weighted by $P_{F}\sim \lmax^{-3}\sim N^{-1}$ for a set of constant density systems.
The Fourier moves are therefore subdominant to the wavelet cost, and the previous $N\ln N$ scaling remains, as observed in Fig.~\ref{fig: Cost semi-dilute}.
Similarly the fractal scaling of $N$ is unchanged if $\lmax$ continues to scale at least as fast as $N^{1/3}$, which is slower than the $N^{\nu}$ scaling of the polymer size.

Unlike the Ewald sum in BD simulations, which splits the workload so that the cost of the Fourier space calculation is comparable to the position space part,\cite{Jain2012BD} there is no requirement for the total cost of WMCD Fourier moves to be comparable to the cost of wavelet moves.
Indeed Fig.s~\ref{fig: Cost semi-dilute} and \ref{fig: Cost dilute} both show the Fourier moves can contribute only a small, if not negligible fraction of the total cost.
$\lmax$ can be tuned to give comparable costs, but since Fourier moves replace wavelets on a 1-to-1 basis regardless of the value of $\lmax$ and they always carry the cost of moving every particle, this is not optimal.
Rather it is optimal to take $\lmax =L/2$, or as close as the algorithm's implementation allows.

\section{Algorithm Validation}
\label{sec: Validation}

\subsection{Chain relaxation}
\label{sec: Chain relaxation}

\begin{figure}
\includegraphics{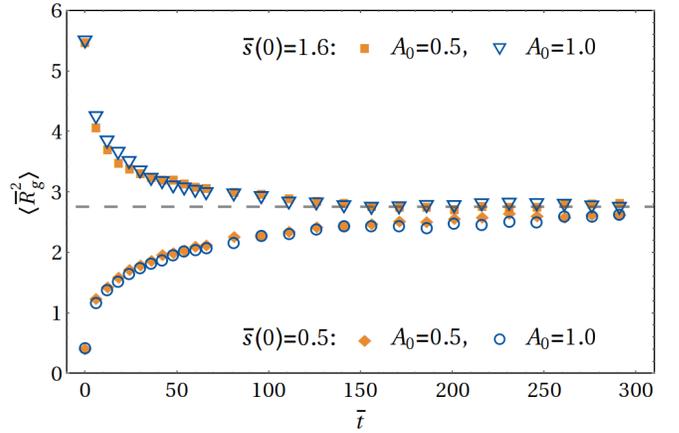}
\centering
\caption{Relaxation of an isolated polymer with $N_{b}=10$ and WCA and FENE potentials, for different $A_{0}$ and initial bond lengths, $\bar{s}(0)$.
The dashed line shows the value for the same system in Ref.~\citenum{Jain2012BD}.}
\label{fig: Relaxation}
\end{figure}

Relaxation of isolated chains of length $N_{b}=10$ beads with initially stretched or compressed bonds is seen in Fig.~\ref{fig: Relaxation}, with mean-square radius of gyration defined as
\begin{equation}
\left\langle R_{g}^2 \right\rangle  =  
	\dfrac{1}{2N_{b}^2} \sum\limits_{i=1}^{N_{b}}  \sum\limits_{j=1}^{N_{b}}
	\left\langle r_{ij}^{2} \right\rangle.
\label{eq: Radius of gyration}
\end{equation}

Both stretched and compressed chains relax to the same value as obtained for this system in Ref.~\citenum{Jain2012BD}, confirming the algorithm equilibrates correctly, and therefore that our $\lambda$ and $k$ recycling schemes have not violated detailed balance.
The figure also shows the same physical relaxation time for the different values of $A_{0}$, verifying the $A_{0}$-dependence in Eq.~\eqref{eq: dt w+F}.

\subsection{Chain diffusivity}
\label{sec: Chain diffusivity}

\begin{figure}
\includegraphics{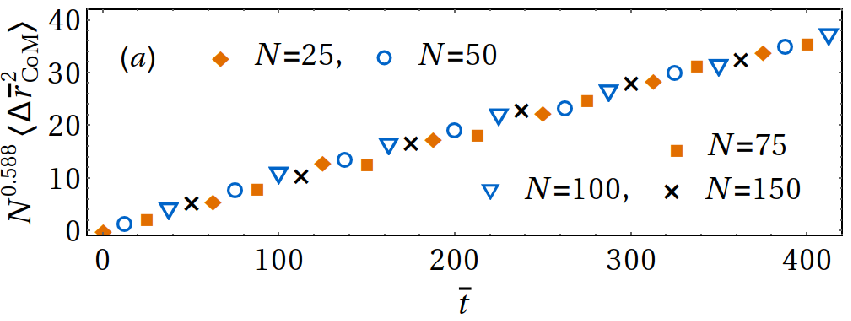}
\includegraphics{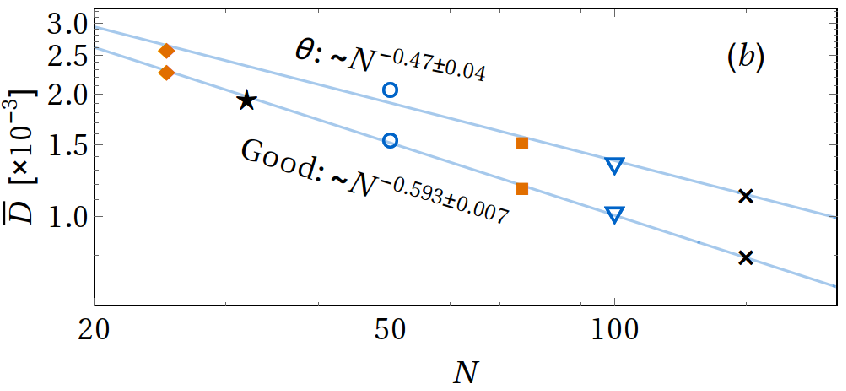}
\centering
\caption{(a): Mean centre of mass displacement squared for isolated chains in a good solvent, scaled to show the theoretical collapse of the data.
Every fifth data point has been used from each data set to show this clearly.
(b): Measured diffusion constants plotted against chain length for chains with identical mean bond lengths in good and $\theta$ solvents.
$A_{0}=0.5$ was used for all WMCD data while the star marks the diffusivity measured in good solvent in Ref.~\citenum{Pham2009LBvsBD}.}
\label{fig: Diffusivity}
\end{figure}

Diffusion of the centre of mass of chains of different lengths is shown in Fig.~\ref{fig: Diffusivity}(a), where the correct linear relationship with time is seen.
Note that we are ignoring the difference between short and long time diffusion in the analysis as this basic data is not precise enough to distinguish the few percent between them.\cite{Liu2003Dshort_vs_Dlong}
Data at that accuracy is left for future work.

Fig.~\ref{fig: Diffusivity}(b) shows measured centre of mass diffusion constants, $D = \left\langle \Delta r_{\mathrm{CoM}}^{2} \right\rangle / 6 t$, for chains in both good and $\theta$ solvents, with the former using the same data as in Fig.~\ref{fig: Diffusivity}(a).
For the $\theta$ solvent the WCA potential was turned off and the FENE potential was replaced with the harmonic bond potential $(3/2)k_{\mathrm{FENE}}(r_{ij} - r_{0})^{2}$, in which $\bar{r}_{0}=0.9691$ was used to match the mean bond length in the good solvent.
These sets of data show quantitatively that simulations in both types of solvent lead to expected scaling with $N$ within the error bounds on the exponent.\cite{Doi1988theory}
The absolute values in good solvent also agree with previous work, with the value at $N=32$ (star) provided by Ref.~\citenum{Pham2009LBvsBD} lying on the fit of our data.
No previous data for $\theta$ solvents with these parameters are available for comparison.

This scaling seen in Fig.~\ref{fig: Diffusivity}(b) is usually considered asymptotic, requiring long chains to observe.
To see why this is not the case with our data we write the chain diffusivity as per Ref.~\citenum{Dunweg2002hydro_radius},
\begin{eqnarray}
D & = & \frac{k_{B}T}{6\pi\eta a} \left[ N^{-1} 
	+ a \left\langle R_{\mathrm{HI}}^{-1} \right\rangle  \right]
\nonumber \\  & = &
\frac{k_{B}T}{6\pi\eta a} \left[ N^{-1} 
	+ a \left( A N^{-\nu} - B N^{-1} + \cdots \right) \right],
\label{eq: D series}
\end{eqnarray}
which has split $D$ into terms expressing the sum of hydrodynamic radii of individual monomers and the contribution from the HIs between the monomers.

Using $B=4.036 /s$, found for Gaussian chains with root mean square bond length $s$,\cite{Dunweg2002hydro_radius} as an rough estimate for our systems, and inputting our values of $s\approx 1$ and $a\approx 0.3$, we find $a B \approx 1.2$.
This leads to significant cancellation of the $N^{-1}$ terms, while $aA\approx 1.1$ so the $N^{-\nu}$ term dominates and scaling is expected at small $N$.
Note that the common practice of neglecting the first term in Eq.~\eqref{eq: D series} would hide this result.

For polymers in good solvent there are additional `non-analytic' terms that come from excluded volume interactions.\cite{Dunweg2002hydro_radius}
Nonetheless the qualitative result is the same when we use coefficients fitted in Ref.~\citenum{Dunweg2002hydro_radius}.
While we do not expect the coefficients used here to apply exactly for our systems, it is the near cancellation of the $N^{-1}$ terms that we highlight as explaining the scaling observed, and that we do expect to apply.

\subsection{Finite box effect on self-diffusion}
\label{sec: Finite box effect}

The periodic algorithm should lower the diffusivity because of the extra HIs with periodic images.
The simplest check for this is to look at the trace of the mobility tensor in the limit $r_{ij}\rightarrow 0$, which has been calculated to leading order to vary with box size as \cite{Dunweg1993box_size_dependence}
\begin{equation}
\mathrm{Tr}[\Oseen^{P} (0;\lmin) ] = \mathrm{Tr}[\Oseen^{\infty} (0;\lmin) ]
	\left( 1 - 2.837 \frac{a}{L} \right).
\label{eq: 1/L dependence}
\end{equation}

\begin{figure}
\includegraphics{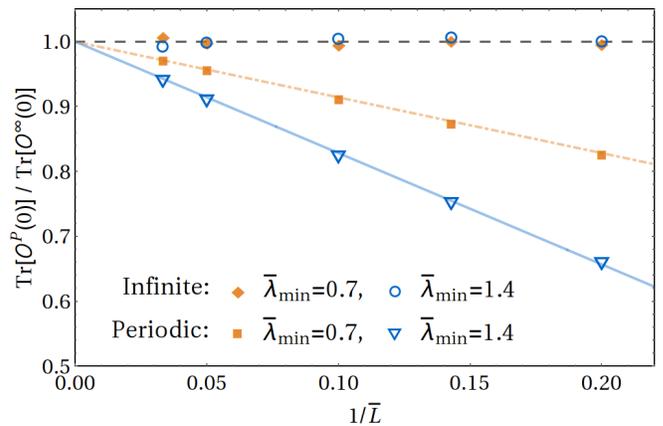}
\caption{Simulated traces of self-diffusion tensors, scaled by the mean trace of infinite systems for each $\lmin$, in simulation boxes with $\bar{L}$ between 5 and 30.
The solid and dot-dashed lines were calculated using Eq.~\eqref{eq: 1/L dependence}.
$A_{0}=0.5$ was used for this data.}
\label{fig: 1/L dependence}
\end{figure}

By generating correlation curves for $\left\langle \delta r_{i}^{2} \right\rangle$ with non-interacting particles, similar to Fig.~\ref{fig: Correlation plots}, and comparing the unscaled values at $r_{ij}\rightarrow 0$, we have confirmed this behaviour as well as the $L$-independence of the infinite algorithm (see Fig.~\ref{fig: 1/L dependence}).
This figure also confirms that Eq.~\eqref{eq: dt w+F} is indeed appropriate for periodic systems as the measured diffusivities are inversely proportional to how we scale time.
Finally, it provides further verification of Eq.~\eqref{eq: lambda_a} as this determines the value of $a$ and hence the gradient of the lines via Eq.~\eqref{eq: 1/L dependence}.

\section{Summary and Outlook}

In this paper we have detailed the Wavelet Monte Carlo dynamics algorithm, using wavelet and plane wave moves to produce hydrodynamics without explicit decomposition of the mobility tensor.
Our algorithm has been shown to closely approximate the Rotne-Prager tensor by starting from the Oseen tensor and removing small radii moves, which in turn provides an explicit particle radius.
The distribution of additional Fourier moves, meanwhile, can be chosen to simulate either a periodic or infinite system with the main wavelet part of the algorithm left unchanged.

It has been shown that the computational cost of WMCD scales well to large systems, going as $N\ln N$ for homogeneous systems with fixed particle density and as $N$ for fractal systems.
Both of these have a very competitive prefactor, making it a promising simulation technique for extending the reach of soft-matter simulations.

The algorithm has been tested to show correct hydrodynamic correlations are produced, while the simulated behaviour of isolated polymer chains has been shown to agree with previous work.

For the case of homogeneous systems,  our wavelet algorithm turns out to be fortuitously well balanced across lengthscales according to three distinct criteria.  
First we chose the move amplitudes (Eq.~\eqref{eq: Wavelet amplitude}) such that their elastic energy is expected to be of order $k_{B}T$ leading to move acceptance independent of wavelet radius $\lambda$.  
Secondly we then found that this leads to computational cost ~$\int d\lambda/\lambda$  uniformly distributed over lengthscales, leading to the logarithmic dependence in the overall cost (Eq.~\eqref{eq: Cost wavelet homogeneous}).  
Thirdly the timescale for the algorithm to build the move correlations across a distance $r$, which come predominantly from wavelet radii $\lambda \sim r$, turns out to be independent of $\lambda$.  
This last result can be seen from the way the squared move amplitudes have $\lambda$-dependence matching the $r$-dependence of the cross mobility (Eq.~\eqref{eq: Oseen tensor}), and it means that the timescale to build up long ranged correlations is of the same order as the timescale for particles to expect to experience a local (small wavelet) move.    
It is easily checked that the $d$-dimensional sketch of the method given in section \ref{sec: Method Sketch} leads to all the same coincidences.  
The case where the coincidences do unravel is when the distribution of particles is fractal, as with dilute polymer and locally in semidilute polymer:  then the longer wavelet moves are allowed greater amplitude leading to computational cost dominated by small wavelets.  
There is correspondingly longer build-up time for long ranged correlations, but this is still fast compared to the corresponding polymer relaxation modes. 

With the core set out, we plan to develop the algorithm further to include solvent flows.
We envision the procedure with this modification to be essentially the same as presented in this article interspersed with flow-generating moves.
So long as the flow moves, however they are implemented, have small enough amplitudes to keep close, interacting particles near equilibrium, even with large scale separations along the chain moved out of equilibrium, we expect the energy estimates in this article to remain valid.
Consequently we do not anticipate any change to how we handle the existing moves.

On the computational side, there is scope to improve the current algorithm in several areas.
Foremost we can improve the accuracy and reduce the variation in move rejection rate by switching from Metropolis to Glauber move acceptance, and also to smart MC in which moves are biased according to the force conjugate to them.  
There is also scope for optimisation of choices which already exist, notably the choice of mother wavelet is yet to be explored.  
The precise way in which periodic boundary conditions are enforced on the hydrodynamic propagator is another deeper choice where we have followed that of prior work, but it is a choice: for example we could have chosen to keep only the wavelet moves.  

A range of coding improvements also stand to be made.  
The computational cost could be reduced by tracking particle neighbour lists rather than re-computing them at each move, and computational speed could be increased by a combination of parallel execution of multiple non-overlapping wavelet moves and parallelised execution of single large wavelet moves.

\begin{acknowledgments}
This work has been supported by the Engineering and Physical Sciences Research Council (EPSRC), grant number EP/L505110/1 (OTD), and the Monash-Warwick Alliance (RCB).
The authors acknowledge stimulating discussions with J R Prakash and are also grateful to the remaining organisers of ``Hydrodynamic Fluctuations in Soft-Matter Simulations" CECAM/Monash workshop, Prato (February 2016) wherein many helpful discussions were had.
\end{acknowledgments}

\begin{appendices}

\appendix
\section{Showing the wavelet representation gives the Oseen tensor}
\label{sec: Wavelet Derivation}

Here we evaluate the integral in Eq.~\eqref{eq: Wavelet representation Oseen} to show that this reduces to a form exactly proportional to the Oseen tensor. 
We begin by taking the Fourier transform with respect to $\textbf{r}_{i}-\textbf{r}_{j}$, leading to 
\begin{eqnarray}
\tilde{\bm{g}}(\textbf{k}) & = & \int
	d^{3}\textbf{r} e^{-ik\cdot (\textbf{r}_{i}-\textbf{r}_{j})}
	\bm{g}(\textbf{r}_{i}-\textbf{r}_{j})  
\nonumber \\
	& = & \mathcal{N}_{\bm{g}} \int
	d^{2}\phat\frac{d\lambda}{\lambda}\lambda^{2}
	\tilde{\textbf{w}}\left(\lambda \textbf{k},\phat\right)
	\otimes
	\tilde{\textbf{w}}\left(-\lambda \textbf{k},\phat\right),
\end{eqnarray}
where $\tilde{\textbf{w}}(\textbf{k},\phat)=\int d^{3}\textbf{x} e^{-i\textbf{k}\cdot\textbf{x}} \textbf{w}(\textbf{x},\phat)$ is the Fourier transform of the mother wavelet at fixed $\phat$. 

We next integrate over the wavelet polarisation to obtain
\begin{eqnarray}
\tilde{\bm{g}}(\textbf{k}) & = &
	4\pi \mathcal{N}_{\bm{g}} \int d\lambda\, \lambda\,\bm{W}(\lambda \textbf{k}); \\
\bm{W}(\textbf{k}) & = & \frac{1}{4\pi}\int
	d^{2}\phat\,
	\tilde{\textbf{w}}
		\left(\textbf{k},\phat\right)
	\otimes
	\tilde{\textbf{w}}
		\left(-\textbf{k},\phat\right)
\nonumber \\
	& = & (\textbf{I}-\hat{\textbf{k}}\otimes\hat{\textbf{k}}) f(k) ,
\end{eqnarray}
which has the explicit tensor structure shown because the constraint $\nabla\cdot\textbf{w}=0$ leads to $\textbf{k}\cdot\bm{W}=0$.
The amplitude factor is set by 
\begin{equation}
2f(k)=\mathrm{Tr}\left[\bm{W}(\textbf{k})\right] = \frac{1}{4\pi} \int d^{2}\phat\, \tilde{\textbf{w}}\left(\textbf{k},\phat\right)
\cdot \tilde{\textbf{w}}\left(-\textbf{k},\phat\right).
\end{equation}
Reassembling all this and making the substitution $q=\lambda k$ leads to 
\begin{equation}
\tilde{\bm{g}}(\textbf{k})= 4\pi \mathcal{N}_{\bm{g}}
	k^{-2} (\textbf{I}-\hat{\textbf{k}}\otimes\hat{\textbf{k}})
	\!\!\! \int\limits_{k\lmin}^{k\lmax} \!\!\! dq\, q f(q).
\label{eq: g tensor wavelet FT}
\end{equation}
For $\lmin\rightarrow 0$ and $\lmax\rightarrow \infty$
this exactly matches the fourier transform of the Oseen tensor, 
$(1/\eta k^{2})(\textbf{I}- \hat{\textbf{k}}\otimes\hat{\textbf{k}})$, upon choosing the normalisation $\mathcal{N}_{\bm{g}}^{-1}=4\pi\eta\int_{0}^{\infty} dq\, q f(q)$.

\section{Simplifying the wavelet strain energy integral}
\label{sec: Simplifying strain integral}

The strain energy in Eq.~\eqref{eq: Wavelet strain energy definition} can be simplified by changing integration variable to $\textbf{x} = \textbf{r}/\lambda$, whereupon all the $\lambda$-dependence in the integral factors out via $d^{3}\textbf{r} = \lambda^{3} d^{3}\textbf{x}$ and $\nabla =\lambda^{-1}\delx$, with $\delx$ the gradient with respect to $\textbf{x}$.
This leaves the integral as a parameter independent number so that $\DUest \propto n/\lambda^{2}$ for \textit{all} mother wavelets. 

The integral itself reduces to
\begin{equation}
A_{w}^{2} \int d^{3}\textbf{x} \left( \delx \textbf{w}:\delx \textbf{w} + 
	\delx \textbf{w}:(\delx \textbf{w})^{T} \right),
\end{equation}
in which the left term is readily seen to integrate to 0 by integrating by parts and then using the divergencelessness of our wavelets (Eq.~\eqref{eq: wavelet divergencelessness}) and that they take the value 0 at their boundary (section \ref{sec: Choice of Wavelet}).
The integral over the remaining term can be written in terms of the wavelet's Fourier transform as
\begin{equation}
\frac{A_{w}^{2}}{(2\pi)^{3}} \int d^{3}\textbf{k}\, k^{2}
	\tilde{\textbf{w}}(\textbf{k})\cdot \tilde{\textbf{w}}(-\textbf{k}).
\end{equation}
We then substitute in the Fourier transform of the wavelet type in Eq.~\eqref{eq: General wavelet}, $\tilde{\textbf{w}}(\textbf{k}) = i(\hat{\textbf{p}}\times \textbf{k})\tilde{\phi}(|\textbf{k}|) $, and use $(\hat{\textbf{p}}\times \textbf{k})\cdot (\hat{\textbf{p}}\times \textbf{k}) = k^{2}\sin^{2}\theta$ to reach the final form of
\begin{equation}
\frac{8\pi A_{w}^{2}}{3(2\pi)^{3}}\int\limits_{0}^{\infty} \! dk \, k^{6} \tilde{\phi}(k)^{2}.
\end{equation}

\section{Derivation of $\Oseen_{w}(r;\lambda',\infty)$}
\label{sec: Derivation of O'}

The inverse Fourier transform of Eq.~\eqref{eq: Oseen_w FT}, with the $q$ integral from $k\lambda'$ to $\infty$, is:
\begin{multline}
\Oseen_{w}(r;\lambda',\infty) = \\
	\frac{A_{0}^{2}}{6\pi^{2}}
	\int\limits_{\mathrm{all}\ \textbf{k}} d^{3}\textbf{k}\,
	e^{i\textbf{k}\cdot\textbf{r}}
	k^{-2} \left(\textbf{I} - \hat{\textbf{k}}\otimes\hat{\textbf{k}}\right)
		\int\limits_{k\lambda'}^{\infty} dq\,q^{3}\tilde{\phi}(q)^{2} .
\end{multline}

First, we commute the $q$ and $k$ integrals, which requires a change in the limits.
To be able to perform the integral we also note that $\hat{\textbf{k}} \otimes\hat{\textbf{k}} e^{i \textbf{k}\cdot \textbf{r}} = -\nabla \otimes\nabla k^{-2}e^{i \textbf{k}\cdot \textbf{r}}$.
What remains is an inverse FT of a radial function, so the angular integrals over the exponential give the usual result of $4\pi \sin(kr)/kr$, so that we have
\begin{multline}
\Oseen_{w}(r;\lambda',\infty) = \\
	\frac{2 A_{0}^{2}}{3\pi}
	\int\limits_{0}^{\infty} dq\,q^{3}\tilde{\phi}(q)^{2}
	\int\limits_{0}^{q/\lambda'} dk\,
	\left(\textbf{I} + \nabla \otimes\nabla k^{-2}\right)
	\frac{\sin{kr}}{kr}.
\end{multline}
We now focus on performing the $k$ integral.

The \textbf{I} term integrates directly to $\mathrm{Si}(qr/\lambda')\,\textbf{I}/r$.
The other term appears to diverge at the $k\rightarrow 0$ limit, but it can be split up into an integral with limits $(0,\infty)$ minus a finite one over $(q/\lambda',\infty)$.
Although the first of these still appears to diverge, it is exactly what would have been integrated if we had the full Oseen tensor.
It is therefore known to be $(\pi/4r)(\hat{\textbf{r}} \otimes\hat{\textbf{r}} - \textbf{I})$. \cite{Doi1988theory}

Hence we are only left with $- \nabla\otimes\nabla $ acting on
\begin{multline}
\frac{1}{r}
	\int\limits_{q/\lambda'}^{\infty}
	dk\, k^{-3}\sin(kr) = \\
 \frac{\lambda'}{2q}
	\left[ \cos Q + Q^{-1}\sin Q + 
		Q\left(\mathrm{Si}\,(Q) - \frac{\pi}{2}\right) \right],
\end{multline}
with $Q=qr/\lambda'$.
Performing the two derivatives obtains the form
\begin{multline}
\left(\mathrm{Si}\,(Q) - \frac{\pi}{2}\right)\frac{1}{2r}
	\left(\hat{\textbf{r}}\otimes\hat{\textbf{r}} - 
		\textbf{I}\right) \\
	+
	\dfrac{\sin Q - Q\cos Q}{Q^{2}}\frac{1}{2r}
	\left(\textbf{I}-
	3\hat{\textbf{r}}\otimes\hat{\textbf{r}}\right).
\end{multline}
The $\pi /2$ term cancels exactly with the full Oseen term above, while the remaining terms plus the contribution from the first $\textbf{I}$ term give the final result
\begin{multline}
\Oseen_{w}(\textbf{r};\lambda',\infty) = \frac{A_{0}^{2}}{3\pi r}
	\int\limits_{0}^{\infty} dq\,q^{3}\tilde{\phi}(q)^2 
	\bigg[ \left(\textbf{I} + \hat{\textbf{r}}\otimes\hat{\textbf{r}}\right)
		\mathrm{Si}\,(Q) \\
		+ \left(\textbf{I} - 3\hat{\textbf{r}}\otimes\hat{\textbf{r}}\right)
		\dfrac{\sin Q - Q \cos Q}{Q^2}
	\bigg].
\end{multline}

Note that this calculation could have started from Eq.~\eqref{eq: g tensor wavelet FT} for a more general wavelet. In this case, all that need changing are $q^{2}\tilde{\phi}(q)^{2}/3 \rightarrow f(q)$ and $\mathcal{N}_{\bm{g}} \rightarrow A_{0}^{2} $.

\section{Details of generating move parameters}
\label{sec: Detailed move generation}

In this section we detail how the calculations of parameter distributions in this article translate to generating the Monte Carlo moves in a WMCD simulation.
We denote numbers generated by a pseudo random number generator by {\tt rand}, which is always uniform over the range indicated.
Where multiple numbers are needed for a single quantity, subscripts are used to distinguish them.

Before any parameters are generated, the move must first choose either a wavelet or a plane wave.
For this we generate ${\tt rand} \in (0,1]$ and check this against $P_{F}$, as determined by Eq.s~\eqref{eq: P(Fourier) infinite} and \eqref{eq: P(Fourier)}.
If ${\tt rand} \leqslant P_{F}$ we generate a plane wave, else we generate a wavelet.

\subsection*{Generating a wavelet}

\paragraph*{Choosing $\lambda$:}
To achieve the distribution over $\lambda$ in Eq.~\eqref{eq: PDF for lambda} we generate ${\tt rand}\in [0,1]$ and use the inverse transform method to convert this to $\lambda\in [\lmin,\lmax]$ with
\begin{equation}
\lambda = \lmin \left( 1 + {\tt rand} \left[(\lmin /\lmax)^{3} - 1\right]\right)^{-1/3}.
\end{equation}

\paragraph*{Choosing \textbf{b}:}
As described in section \ref{sec: Centring of Wavelets} we first pick a particle by generating $i = {\tt rand} \in \left\lbrace 1,2,3,...,N\right\rbrace$ and then pick \textbf{b} uniformly in a sphere of radius $\lambda$ around $\textbf{r}_{i}$.
To do this we generate a vector uniformly distributed inside the unit sphere by repeatedly generating $\textbf{V} = ({\tt rand}_{x},{\tt rand}_{y},{\tt rand}_{z} )$ with each ${\tt rand}\in [-1,1]$ until $\textbf{V}\cdot\textbf{V} \leqslant 1$.
Then \textbf{b} is given by
\begin{equation}
\textbf{b} = \textbf{r}_{i} + \lambda \textbf{V}.
\end{equation}

\paragraph*{Choosing $\phat$:}
$\phat$ is an isotropically distributed unit vector, for which we generate \textbf{V} as above and take $\phat = \textbf{V}/\sqrt{\textbf{V}\cdot\textbf{V}}$.
\\

\paragraph*{Determining $A_{w}$:}
With $\lambda$ and \textbf{b} chosen we can simply count how many particles, or a periodic images if appropriate, are centred inside it to give us the value of $n$.
$A_{w}$ is then determined by Eq.~\eqref{eq: Wavelet amplitude}.
Note that only 1 image should be counted per particle as this is all a single image of the wavelet would contain.

\subsection*{Generating a plane wave}

\paragraph*{Choosing $\textbf{k}$ for an infinite system:}
With infinite systems we can generate the amplitude and direction separately, with the latter generated in the same way as $\phat$ for wavelets, i.e. $\hat{\textbf{k}}= \textbf{V}/\sqrt{\textbf{V}\cdot\textbf{V}}$.

The distribution over the amplitude in Eq.~\eqref{eq: PDF of k} depends on the choice of mother wavelet and its associated cumulative distribution function is not in general analytically invertible. 
Nonetheless we use the inverse transform method numerically using a file containing pre-calculated conversion values from ${\tt rand}\in (0,1]$ to $k$.
Note we interpolate between the necessarily discrete values in the file and exclude ${\tt rand}=0$ because $\mathcal{P}_{k}^{\infty}(0)=0$.
\\

\begin{figure}
\includegraphics{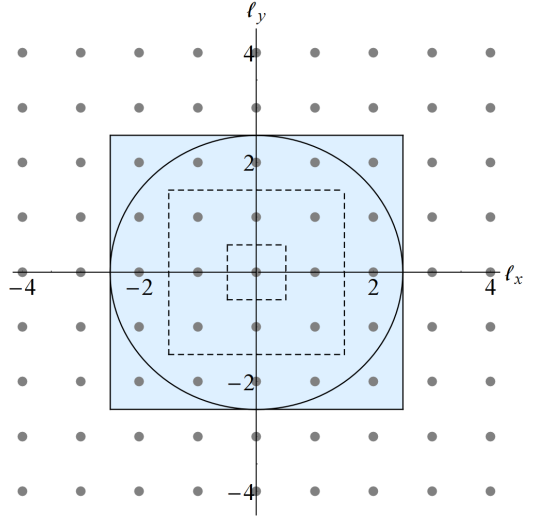}
\caption{2 dimensional diagram of discrete Fourier modes.
The shaded region, here with 2 layers around (0,0) with inner edges marked by dashed lines, contains modes chosen discretely.
The circle indicates the inner edge of the quasi-continuous region.}
\label{fig: discrete k boundary}
\end{figure}

\paragraph*{Choosing $\textbf{k}$ for a periodic system:}
To generate discrete wavevectors we split $\textbf{k}$-space into a cubic discrete region centred on $\textbf{k}$=$\textbf{0}$ and with $\Nlayers$ layers, as shown in Fig.~\ref{fig: discrete k boundary}, and a quasi-continuous region outside.
We choose between these regions by comparison of ${\tt rand}\in (0,1]$ to the sum of probabilities of modes inside the discrete region.

In the discrete region we generate $\textbf{k} = (2\pi /L)({\tt rand}_{x},{\tt rand}_{y},{\tt rand}_{z})$ with each ${\tt rand} \in \lbrace -\Nlayers, -\Nlayers+1, -\Nlayers+2, ... , \Nlayers \rbrace$.
If $({\tt rand} \in (0,1] ) \leqslant P_{\textbf{k}}^{P}(\textbf{k}_{\ell})/P_{\textbf{k}}^{P}(\textbf{0})$, where the denominator serves to increase the acceptance rate without altering the relative probabilities, then we use this mode, else we repeat the process until a mode is accepted.
For the results in this article we have used $\Nlayers=15$, but owing to the rapid $k^{-4}$ decay in probabilities a smaller value can be used while maintaining an accurate distribution.

In the quasi-continuous region we pick modes continuously as per the the infinite case, and then shift them to the nearest discrete mode.
This is accurate when $\mathcal{P}_{k}^{\infty}$ is linear over its surrounding Wigner-Seitz cell, as is ensured locally by the Taylor series.
Since the distribution now starts at $k=(2\pi/L)(\Nlayers+0.5)$ rather than 0, a different conversion table to the one used in the infinite algorithm must be constructed.
Finally, to not over-represent modes contained in both the discrete and quasi-continuous regions, such as those in the corners of the square in Fig.~\ref{fig: discrete k boundary}, if these modes are landed on in the quasi-continuous region another mode must be selected instead.
\\

\paragraph*{Choosing $\Phi$ and $\hat{\textbf{e}}$:}
The plane wave's phase is simply $\Phi = {\tt rand}\in [0,2\pi)$.
Its orientation, chosen uniformly over the directions perpendicular to $\textbf{k}$, also generates ${\tt rand}\in [0,2\pi)$ as an angle about the $\textbf{k}$-axis.
Geometrical calculations then readily find $\hat{\textbf{e}}$ from {\tt rand} and $\hat{\textbf{k}}$.

\subsection*{Parameter recycling and updating time}

If a move is rejected, we reuse either $\lambda$ or $k$ for wavelets and plane waves respectively to keep their distributions as required for hydrodynamics.
For wavelets and infinite system Fourier moves this just requires using the same $\lambda$ or $k$ as the previous move (note that the same move type is required also), while new values for all other parameters are generated as above.
For periodic Fourier moves the discrete $\textbf{k}$ requires a different approach, and we ensure we keep the same amplitude by simply permuting the components, each with a uniform probability to land in any of the 3 new components, and assign a minus sign to each with a probability of 0.5.

Because the $\lambda$ and $k$ recycling scheme ensures the correct distribution of \textit{accepted} moves, we must update time only after an accepted move, when we add $\delta t$ as given by Eq.~\eqref{eq: dt w+F} to the evolved time.

\end{appendices}

\bibliography{Wavelet_Method_Bib}

\end{document}